# НАЦИОНАЛЬНЫЙ ОТЧЕТ МЕЖДУНАРОДНОЙ АССОЦИАЦИИ ГЕОДЕЗИИ МЕЖДУНАРОДНОГО ГЕДЕЗИЧЕСКОГО И ГЕОФИЗИЧЕСКОГО СОЮЗА 2007-2010

## NATIONAL REPORT FOR THE INTERNATIONAL ASSOCIATION OF GEODESY OF THE INTERNATIONAL UNION OF GEODESY AND GEOPHYSICS 2007–2010

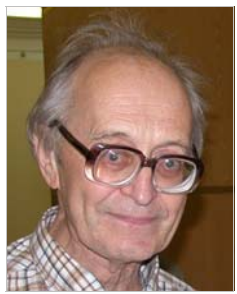
**Боярский Э.А.
Boyarsky E.A.**
Институт Физики Земли им.О.Ю.Шмидта РАН / The Schmidt Institute of Physics of the Earth of Russian Academy of Sciences. e-mail: ernst@ifz.ru

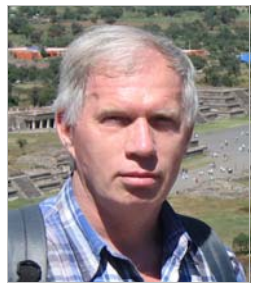
**Витушкин Л.Ф.
Vitushkin L.F.**
ВНИИ метрологии им. Д.И.Менделеева / All-Russian D.I.Mendeleyev Research Institute for Metrology. e-mail: eluar@mail.ru

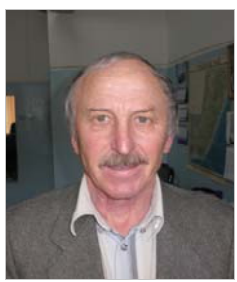
**Герасименко М.Д.
Gerasimenko M.D.**
Институт прикладной математики ДВО РАН / Institute of Applied Mathematics, Far Eastern Branch, Russian Academy of Sciences.
e-mail: mdger@iam.dvo.ru

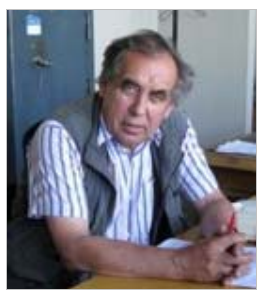
**Демьянов Г.В.
Demianov G.V.**
ЦНИИ геодезии, аэросъемки и картографии / Central research Institute of Geodesy, Aerial Survey and Cartography. e-mail:gleb@geod.ru

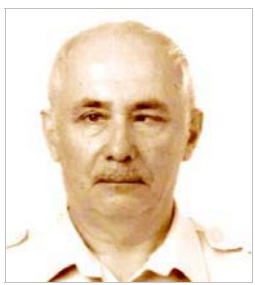
**Кауфман М.Б.
Kaufman M.B.**
ВНИИ физико-технических и радиотехнических измерений / National Research Institute of Physicotechnical and Radio Engineering Measurements of Federal Agency of Technical Regulation and Metrology.
e-mail: mark@imvp.ru

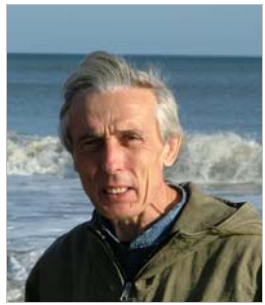
**Кафтан В.И.
Kaftan V.I.**
ЦНИИ геодезии, аэросъемки и картографии / Central research Institute of Geodesy, Aerial Survey and Cartography. e-mail: kaftan@geod.ru

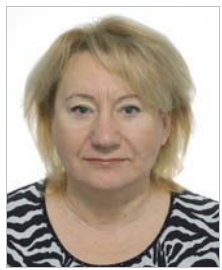
**Мазурова Е.М.
Mazurova E.M.**
Московский государственный университет геодезии и картографии / Moscow State University of Geodesy and Cartography. e-mail: e_mazurova@mail.ru

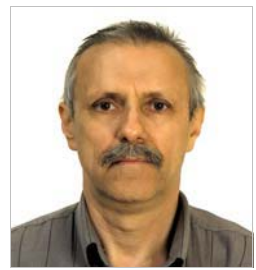
**Малкин З.М.
Malkin Z.M.**
Пулковская обсерватория РАН / Pulkovo Observatory of RAS.
e-mail: malkin@gao.spb.ru

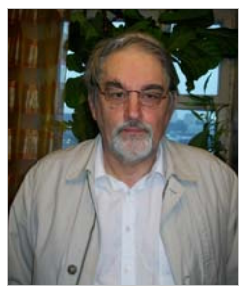
**Молоденский С.М.
Molodenskii S.M.**
Институт Физики Земли РАН / Shmidt Institute of the Physics of the Earth RAS.
e-mail: molodensky2008@rambler.ru

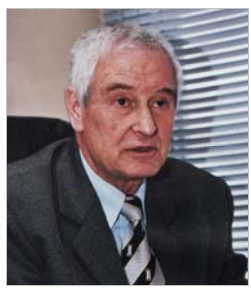
**Нейман Ю.М.
Neyman Yu.M.**
Московский государственный университет геодезии и картографии / Moscow State University of Geodesy and Cartography.
e-mail: yuney@miigaik.ru

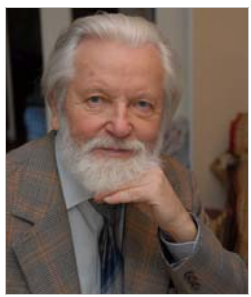
**Певнев А.К.**
**Pevnev A.K.**
Институт Физики Земли РАН / Institute of Earth Physics, Russian Academy of Sciences. e-mail: an.pevnev@yandex.ru

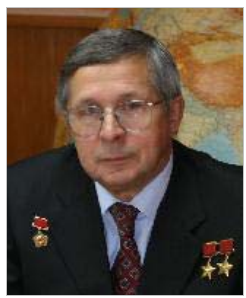
**Савиных В.П.**
**Savinykh V.P.**
Московский государственный университет геодезии и картографии / Moscow State University of Geodesy and Cartography
e-mail: portal@miigaik.ru

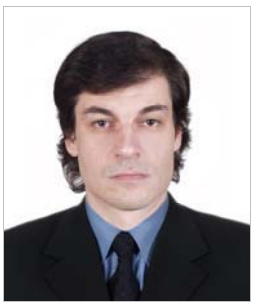
**Стеблов Г.М.**
**Steblov G.M.**
Институт Физики Земли РАН / Institute of Earth Physics, Russian Academy of Sciences.
e-mail: steblov@gps.gsras.ru

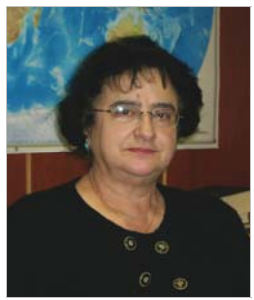
**Татевян С.К.**
**Tatevian S.K.**
Институт Астрономии РАН / Institute of Astronomy of the Russian Academy of Sciences. e-mail: statev@inasan.ru

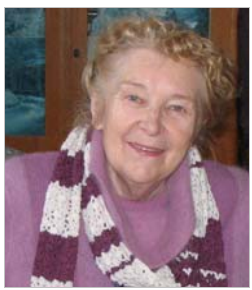
**Толчельникова С.А.**
**Tolchel'nikova S.A.**
Пулковская обсерватория РАН / Pulkovo Observatory of RAS.
e-mail: mchubey@gao.spb.ru

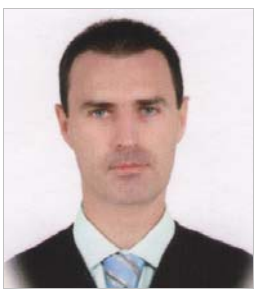
**Шестаков Н.В.**
**Shestakov N.V.**
Дальневосточный федеральный университет, Университет Кванун, г. Сеул, Южная Корея/ / Far Eastern Federal University, Kwangwoon University, Seoul, South Korea. e-mail: nikon@phys.dvgu.ru

**Аннотация.** Национальный отчет, представляемый Международной ассоциации геодезии Международного геодезического и геофизического союза, содержит результаты, достигнутые российскими геодезистами в 2007-2010 гг. В отчете кратко рассмотрены основные результаты исследований и разработок в области геодезии, геодинамики, гравиметрии, в изучении проблем построения и поддержания координатных систем отсчета, изучении фигуры и гравитационного поля Земли, параметров вращения Земли, решения прикладных задач и некоторых других направлений. Период развития геодезии в России с 2007 по 2010 гг. был достаточно сложен, главным образом, из-за непрерывного преобразования государственных геодезических структур и организаций, а также национального образования. Отчет представляет собой обзор основных публикаций и докладов на симпозиумах, конференциях и семинарах. Каждый из разделов отчета включает перечень публикаций за 2007-2010 гг., в том числе подготовленных в соавторстве с зарубежными коллегами. В тексте также отражены некоторые наиболее интересные международные и национальные мероприятия в области геодезии. По ряду объективных причин в отчет помещены не все результаты российских исследований и разработок в области геодезии.
**Ключевые слова:** Геодезия, геофизика, вращение Земли, системы отсчета, геодинамика.

**Abstract.** This report submitted to the International Association of Geodesy (IAG) of the International Union of Geodesy and Geophysics (IUGG) contains results obtained by Russian geodesists in 2007-2010. In the report prepared for the XXV General Assembly of IUGG (Australia, Melbourne, 28 June – 7 July 2011), the results of principal researches in geodesy, geodynamics, gravimetry, in the studies of geodetic reference frame creation and development, Earth's shape and gravity field, Earth's rotation, geodetic theory, its application and some other directions are briefly described. The period from 2007 to 2010 was still difficult for Russian geodesy mainly due to the permanent reformation of state geodetic administration as well as state education structure and organization. The report is organized as a sequence of abstracts of principal publications and presentations for symposia, conferences, workshops etc. Each of the report paragraphs includes a list of scientific papers published in 2007–2010 including those prepared in cooperation of Russian scientists and their colleagues from other countries. Some interesting international and national scientific events are reflected in the text too. For some objective reasons not all results obtained by Russian scientists on the problems of geodesy are included in the report.
**Keywords:** Geodesy, geophysics, earth's rotation, reference systems, Geodynamics.

Scientific edition by Dr.V.P.Savinikh and Dr.V.I.Kaftan

## 1. REFERENCE FRAMES
(Demianov G.V., Kaftan V.I., Mazurova E.M., Tatevian S.K.)

In the frames of the National Program on the improvement of the Fundamental Astro-Geodetic Network (FAGN) a new network of the nine GPS/GLONASS sites are established in Russia. Together with the IGS sites operating till now, there are more than 30 permanent GNSS sites in the country with combined receivers. Several of these sites are collocated with the seismological automatic stations of the IRIS network. The main objective of the Program is definition, realization and maintenance of the unified Reference Frame, focusing on both spatial and vertical components. Every site of this network is tied by precise leveling with the two nearest points of the State Leveling network (in total 400000 km of lines). The repeated absolute gravity measurements at the permanent GPS/GLONASS sites are carried out for estimation of precise vertical components of the coordinates. Several of these sites are already collocated with the existing SLR, VLBI and DORIS stations, and a few additional collocations are planned for further improvement in accuracy and stability of the reference network. Integrating with GGOS project is also considered.

The GPS/GLONASS measurements from all these sites are collected and analyzed by the Analysis center of the Central Research Institute of Geodesy and Cartography, which has been established in cooperation with the BKG. Joint scientific analyses of all space geodetic observations will be carried out in collaboration with the analyses centers of the Russian academy of sciences and Space Agency [Demyanov G.V., Tatevian S.K., 2010].

At present Russian coordinate reference frame (RUREF) consists of the four main geodetic networks of different orders:

- Fundamental astro-geodetic network FAGN (45 stations)
- Precise geodetic network PGN (about 250 points)
- 1st order satellite geodetic network SGN-1 (created in economical developed regions)
- Classical State geodetic network SGN of 1-4 orders and densification (about 500 000 points).

Two highest orders of the RUREF are shown at the Fig.1.

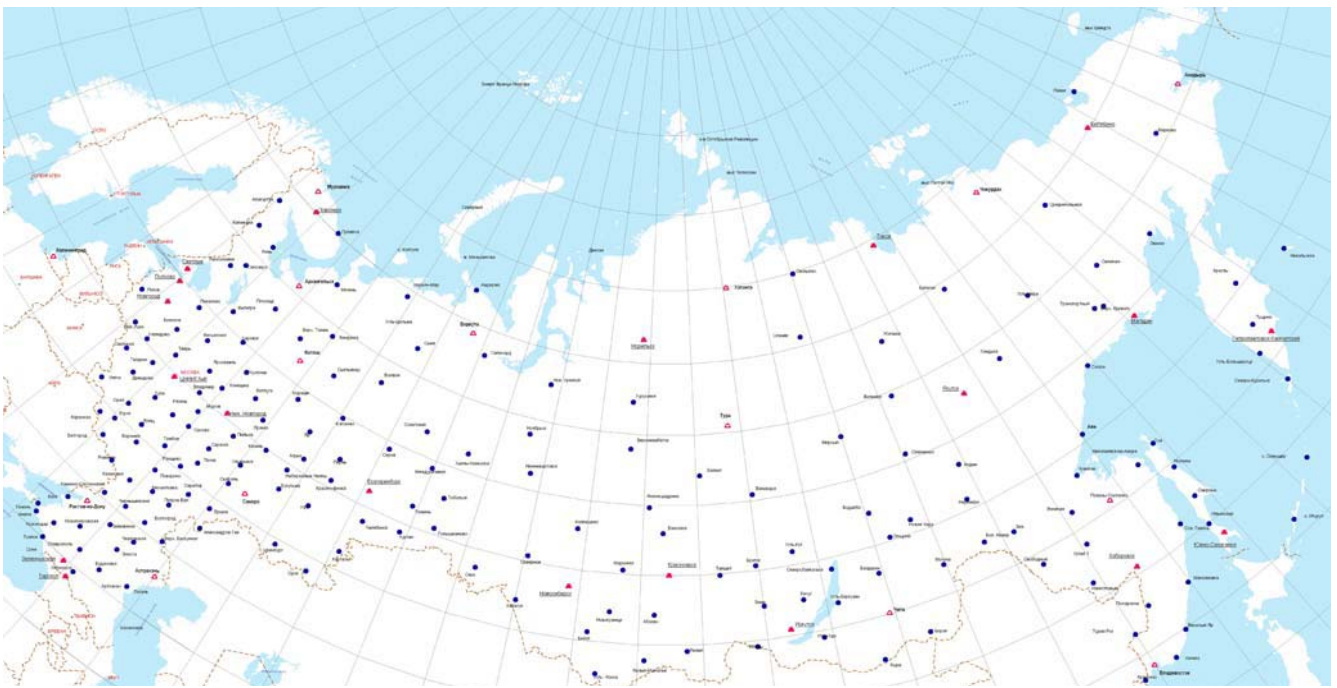

Fig.1. Fundamental astro-geodetic (red trapeziums) and precise geodetic (blue dots) networks of the Russian coordinate reference frame

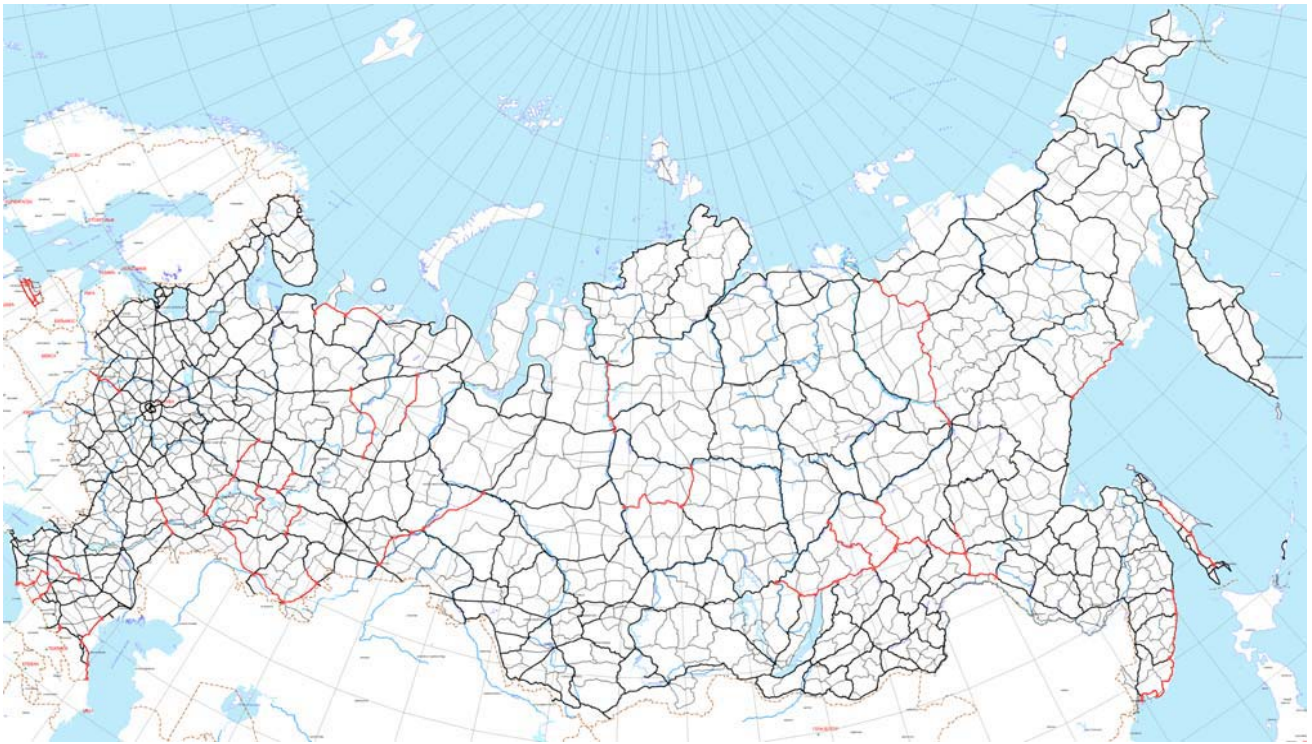

Fig.2. Main Russian Vertical Reference Frame (first order leveling – solid, second order leveling – thin, planned - red)

FAGN consists of 37 permanent GNSS stations and 8 periodically observed satiations. The 25 permanent stations belong to the Federal Service of Registration, Cadastre and Cartography (Rosreestr). Russian Academy of Sciences (RAS) and Federal Agency on Technical Regulation and Metrology (Rosstandart) maintain, 10 respectively, 9 and 4 permanent sites jointly with Rosreestr. The rest 7 permanent sites belong to RAS. Three of them are the VLBI observatories of the Institute of Applied Astronomy of RAS.

All another PGN, SGN-1 and SGN points are in responsibility of Rosreestr.

The main part of PGN sites equipped by special geodetic concrete piers using special GNSS antenna mounts. See example in Fig. 3.

Every FAGN and PGN site includes the main GNSS geodetic center, 2 benchmarks of precise spirit leveling, and 2 ground pillars of the classical SGN of the 1-2 orders tied each other by precise GNSS measurements.

Geodynamical test areas are established in the frame of the common national reference frame and respond to the specifications of PGN-1 or precise permanent local networks.

Central Research Institute of Geodesy, Aerial Survey and Cartography has organized Precise GLONASS Ephemeris Center. Cooperative Virtual Data Analysis Center is working at the base of the Precise GLONASS Ephemeris Center in response of the agreement between Federal Agency of Registration, Cadastre and Cartography (former Roskartography) and Bundesamt für Kartographie und Geodäsie (BKG) [http://84.47.145.86].

Observation and monitoring of global changes require a construction and continuous maintenance of multidisciplinary reference frames. Monitoring of global deformations and mass exchange processes is a main task of IAG's Global Geodetic Observing System. In view of this, global distributed gravimetric networks present the observation component of high/highest importance. The long and successful experience of global space reference frame development allows us to differentiate reference frame and reference system concepts. In this case three main and clear definitions are distinguishing:

- Coordinate systems are abstract mathematical models.
- (Coordinate) reference systems are semiabstract physical models.
- (Coordinate) reference frames are real technological model or network.

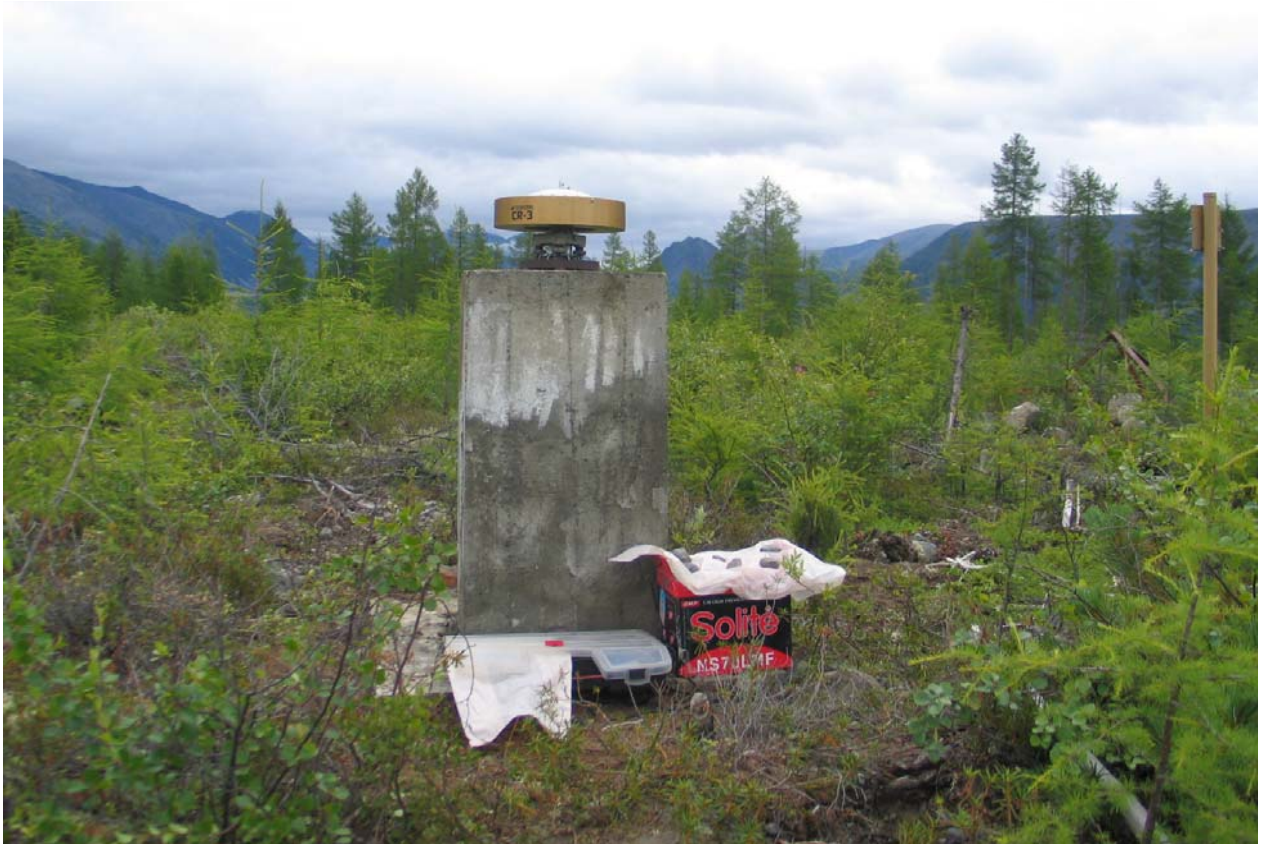

Fig.3 The point of the precise geodetic network

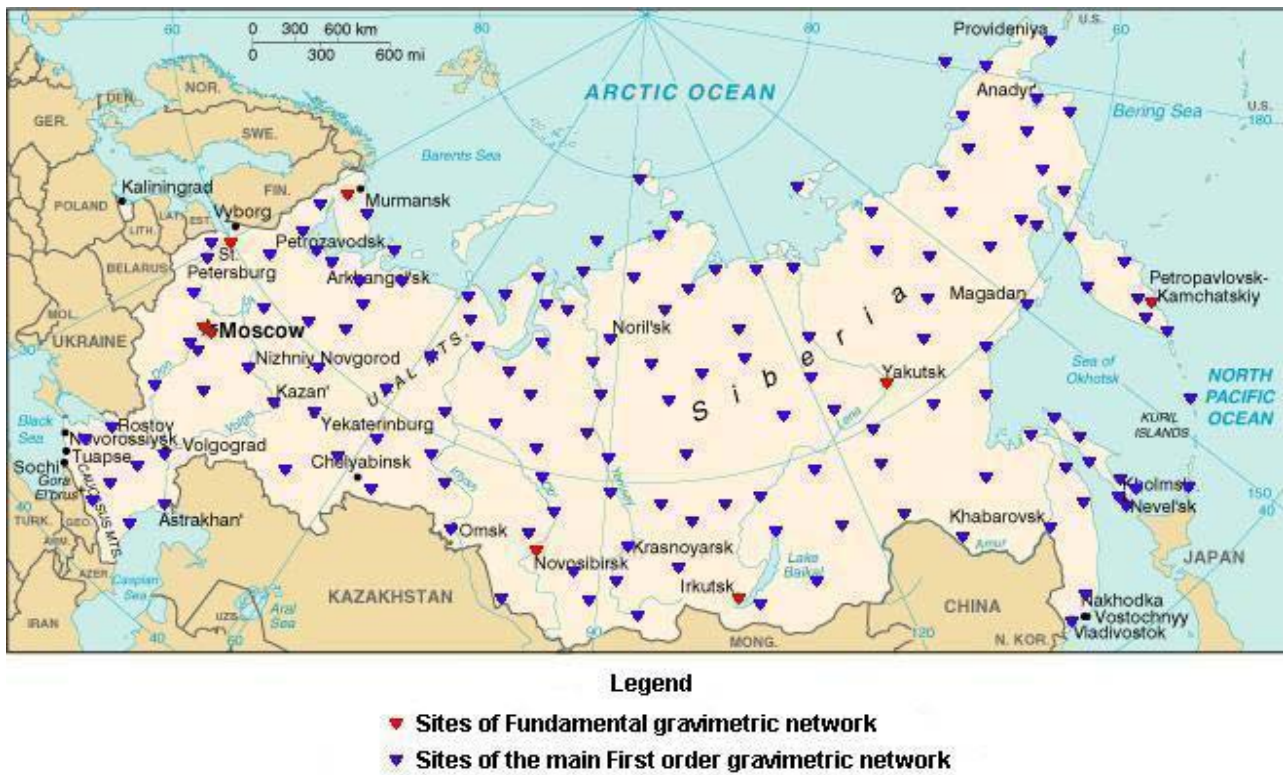

Fig.4. Main gravimetric 1-st order network/reference frame of Russia.

Global gravity monitoring today becomes the task of global space reference frames. It necessitates harmonizing a particular terminology. Today the main global gravimetric network/frame is the International Absolute Gravity Basestation Network (IAGBN). It needs to be closely collocated with sites of the International Terrestrial Reference Frame (ITRF) and modernized to a permanent service by coinciding with superconducting gravimeter network. Therefore it will useful to have more rigorous definitions of gravimetric system and/or gravimetric reference system and gravimetric reference frame.

Several papers are devoted to discussion of the necessity of harmonizing modern and classical definitions in geodetic usage [Kaftan V.I., 2008abc, Kaftan V.I., 2009a].

The condition of the Russian gravimetric reference frame is described in [Gusev N.A., Scheglov S.N., 2010].

The main features of a national reference frame are its complexity and metrological provision. International comparisons of absolute gravity meters are the more important supports of metrological quality of national reference frames.

The series of international comparisons of absolute gravity meters are described in [Makinen J., et all, 2008a & b]. The first comparison was performed in 2004 at the Metsähovi Geodetic Observatory (latitude 60 deg 13 min N, longitude 24 deg 24 min E) of the Finnish Geodetic Institute (FGI). Simultaneous measurements were made between July 6 and July 17 with three gravimeters on three piers (AB, AC, AD). Each gravimeter had two independent setups on each pier. These gravimeters were the FG5 no. 101 and FG5 no. 301 of BKG, and the GBL-P no. 001 of TsNIIGAiK. The FG5 no. 220 of IfE performed a more succinct program on piers AB and AC July 3 to 5. The FG5 no. 221 of the FGI only could perform the full program September 28 to October 8. The possibility of using the record of the superconducting gravimeter GWR T020 at Metsähovi to connect this far-apart (in time) absolute measurement to the others is investigated. The second comparison was performed in October 2005 at the Zvenigorod Observatory (latitude 55 deg 42 min N, longitude 36 deg 46 min E) of the Institute of Astronomy of the Russian Academy of Sciences, on two piers. They were occupied by the GBL-P no. 001 and the FG5-110 of the TsNIIGAiK, and by the FG5-221 of the FGI, with two setups on each pier. The third comparison was performed in June 2007 at two field sites: Pulkovo (latitude 59 deg 46 min N, longitude 30 deg 20 min E) and Lovozero (latitude 67 deg 53 min N, longitude 34 deg 37 min E). Both sites have a single pier. The FG5-110 of the TsNIIGAiK and the FG5-221 of the FGI participated, with two setups at each site. In Metsähovi and Zvenigorod a relative network between the piers were observed at multiple levels using Scintrex CG-5 and LaCoste&Romberg feedback gravimeters. Researchers evaluate the results, analyze the role of the relative measurements, and discuss the methodology of combining several comparisons for the determination of their reference gravity values and of the offsets of the gravimeters.

The fourth comparison was performed in Irkutsk in November 2010 at the site of the FGUP VostSib AGP (East Siberian Division of the Federal Service of Registration, Cadastre and Cartography). See Fig. 5.

Two new Russian gravity meters were compared using four postaments. The FG5 no. 221 of the FGI could not be transferred to the site due to unexpected custom problems. The next gravimeter comparison planned to be performed in 2011 in Moscow.

The comparison was carried out together with the International Meeting on Problems of Geodesy, Geodynamics and Gravimetry at the East Siberian Division of the Federal Service of Registration, Cadastre and Cartography (Fig.5).

Complex character of large scale reference frames requires the accomplishment of special local geodetic ties between points of different techniques. Research on the usage of GPS for VLBI and GPS reference point's collocation are presented in the paper [Kaftan V.I., 2009b]. Preliminary experiments on three Russian VLBI sites of Institute of Applied Astronomy of the RAS showed that the accuracy of a tie has a sub centimeter level. More accurate ties using GPS need to provide the best observation conditions. In a contrast to geodetic electro distance and optical ties, GPS ties have not transformed coordinate errors immanent for the classical approach.

The papers [Mazurova E.M. 2009, Mazurova E.M., Malinnikov V.A., 2010] discuss the history of the GLONASS, modern tendencies and prospects in its development. The main differences between GLONASS and GPS are considered. The role of the systems in global reference frame construction is described.

The history of GLONASS development has been long and difficult. Technical, but mainly political problems of the country have not allowed the system to develop dynamically. Over the last seven years there have been essential changes both in the development of the system and in the field of its application. GLONASS satellites are constantly modernized (GLONASS, GLONASS-M, GLONASS-K, GLONASS-KM). Their satellite life time increases and the number of carrier frequencies also increases, which results in the system higher accuracy.

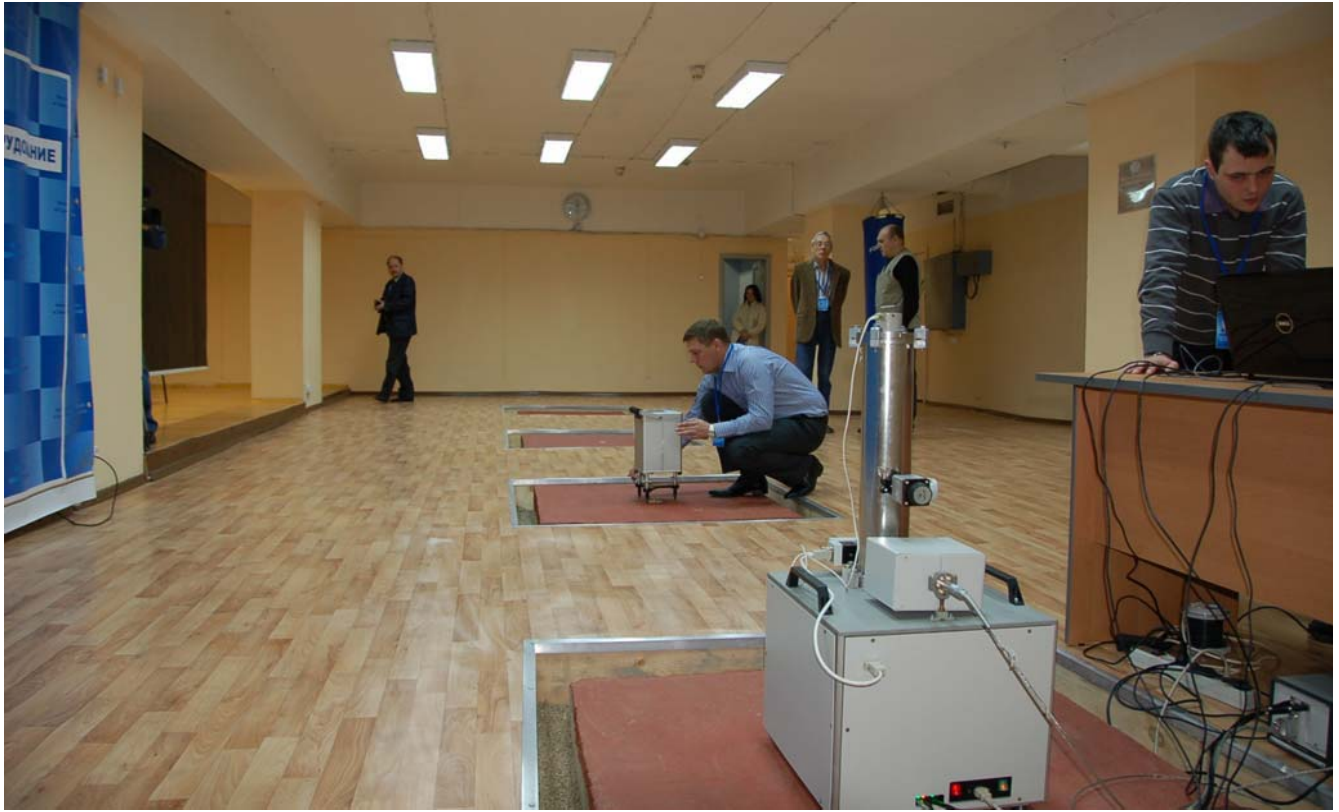

Fig. 5 Comparison of absolute gravity meters at the site of East Siberian Division of the Federal Service of Registration, Cadastre and Cartography

Russia worked out a state geocentric coordinate system PE-90 (PZ-90) to ensure geodetic support for orbital flights and solve navigation problems. Along with the coordinate system PE-90 (PZ-90), Russia is actively using WGS-84. Therefore the paper analyzes the two coordinate systems and shows their relationships. Visibility zones of the GLONASS over Newcastle are given according to the information of the analytical centre of coordinate-time and navigation support of Russia.

The instability of the GLONASS does not allow yet the Russian system to compete on equal footing with the GPS. However, even now, the GLONASS can be used for many purposes not only in Russia, but also abroad.

The International conference "Space geodynamics and global process modeling" (APSG 2008) was held on 22-26 September 2008 in Novosibirsk. Report on the common usage GPS and GLONAS during the Russian Reference Frame development was presented by S.K. Tatevian, S.P.Kuzin [Tatevian S., Kuzin S., 2008].

GLONASS is a key element of the Russian Positioning, Time and Navigation service. The GLONASS state program foresees the full operation capability (24 satellites) was reached by the end of 2010. The second generation of satellites GLONASS-M have been launched with the L2 civil signal, extended lifetime and improved clock stability. GLONASS-K satellites are expected to have the L3 civil signal and Synthetic Aperture Radar function. An extension of the ground control and monitoring network up to 9-12 stations are planning and cooperation with GNSS is 17 foreseen. The combined GPS/GLONASS system is a main technology for development of the fundamental geodetic network and for crust movement studies along the North Eurasian tectonic plate. In combination with GPS, GLONASS can benefit users already now especially in the urban areas. Results of GLONASS data analysis performed at the Institute of Astronomy (Moscow) with the use of GIPSY-OASIS2 software for 16 globally distributed sites show that residuals of the GLONASS solution are almost similar to those from the GPS solution [Kuzin S., Revnivykh S., Tatevyan S., 2007].

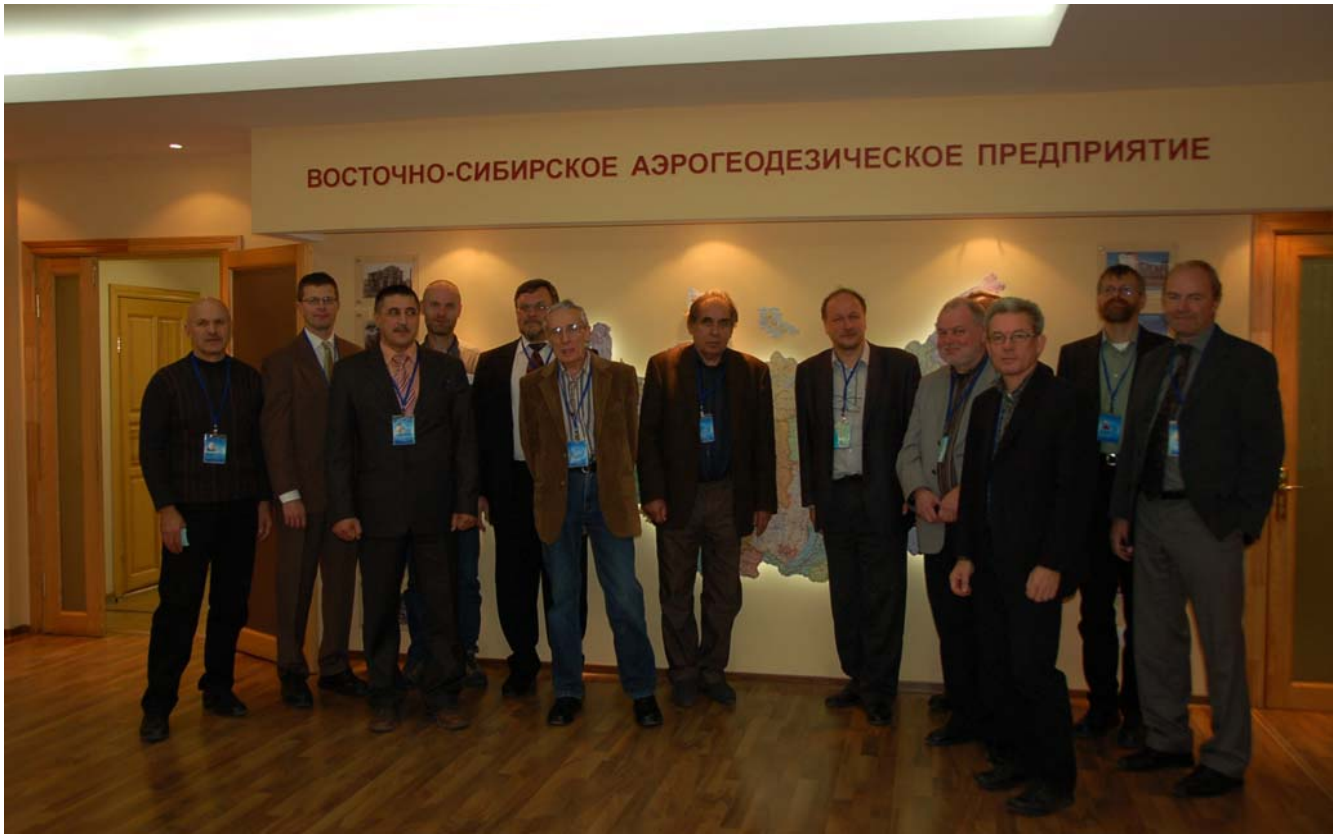

Fig. 6 The participants of the International Meeting on Problems of Geodesy, Geodynamics and Gravimetry

Taking into account recommendations of the International Earth Rotation and Reference Systems Service (IERS) and International DORIS service (IDS), a reprocessing of the DORIS data for the period of 1993.0-2010.0 has been performed aiming to obtain a unified coordinated solution of the IDS analysis centers for the developing of the new version of the Terrestrial Reference Frame - ITRF2008. The weekly solutions of coordinates of all 71 DORIS ground sites and Earth Observations parameters (EOP) have been estimated with the use of new improved satellite surface models, submitted by CNES, and with measurement data of the satellites SPOT2, SPOT3, SPOT4, SPOT5, TOPEX, and ENVISAT. After the transformation of the free-network solution into a well-defined reference frame ITRF2005 (exactly, long-term cumulative IGN solution: ign07d02) weekly coordinates of the sites were estimated with the internal precision at the level of 5-40 mm for majority of the stations.

Russian scientific community pays special attention for historical fragments of the national reference frame. There is a series of works devoted to the "Struve Geodetic Arc" research and public awareness [Kaptüg V., 2008, 2009, 2010; Mazurova E.M., Aleksashina E.V., 2010].

Procedures of space fixation of global coordinate system are proposed in papers [Gerasimenko M.D., Kolomiets A.G., 2008, Kolomiets A.G., 2010]. The authors recommend using the free global network adjustment in a reference to more stable vertical or horizontal site velocities. The approach allows the authors to reject some quality geological models in the process of orientation site velocity vectors.

Russian gravimetric team from Central Research Institute of Geodesy, Aerial Survey and Cartography, and Moscow State University of Geodesy and Cartography, and Vietnam Institute of Geodesy and Cartography (VIGAC) carried out the joint Russian-Vietnam field campaign on gravimetric reference frame modernization in Socialist Republic of Vietnam in 2010-2011. The gravity measurements have performed by new Russian absolute ballistic gravity meter for field operation constructed by the Institute of Automation and Electrometry of the Siberian Branch of the Russian Academy of Science (Fig. 5) and described in [Stus Yu.F., et al., 2010]. The field campaign scenes showed in photos below.

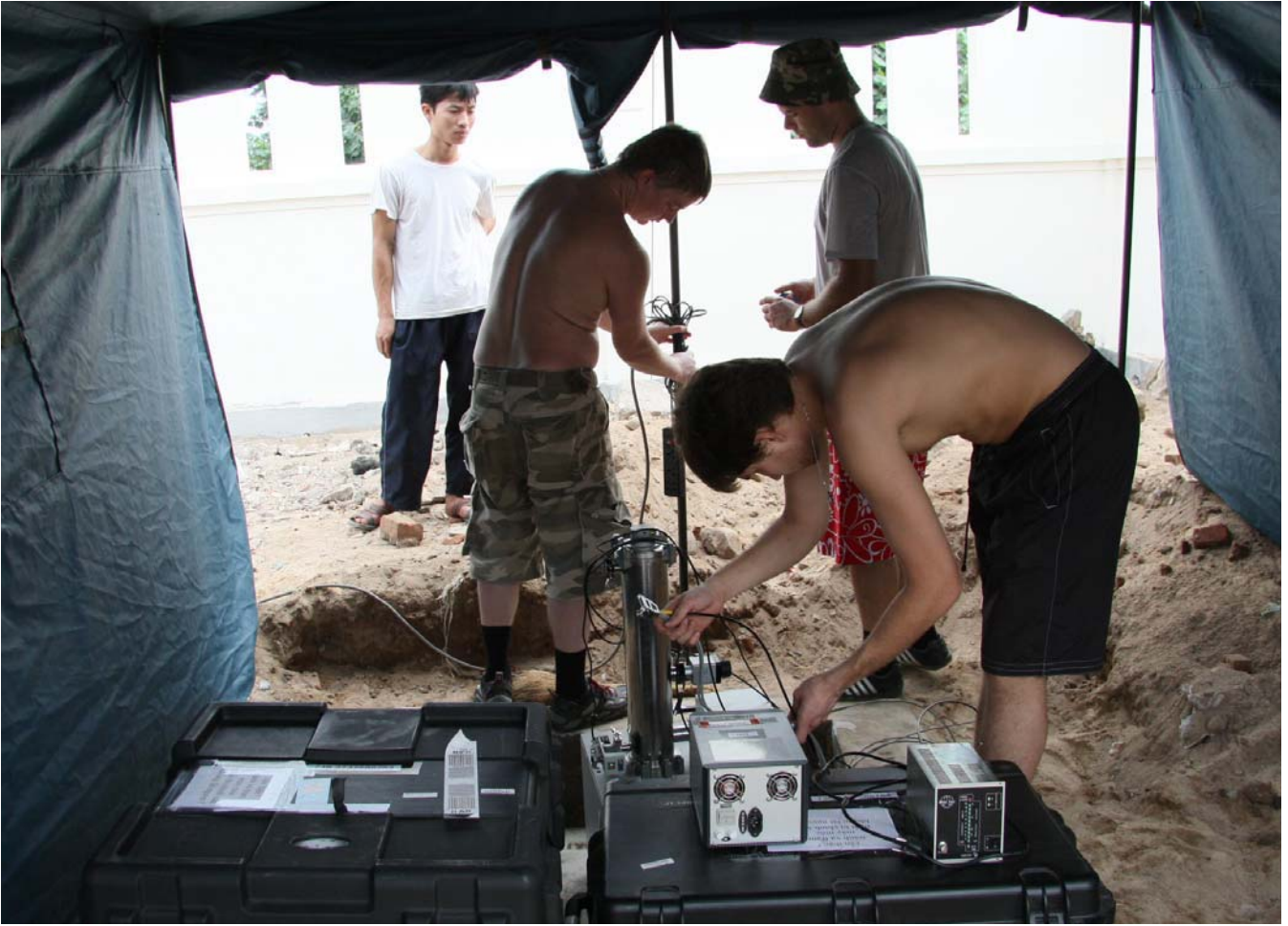

Fig.7 Absolute gravity measurement at a field gravimetric point.

## 2. GRAVITY FIELD
(Boyarsky E.A., Mazurova E.M., Vitushkin L.F)

Questions of calculation of anomaly of height on basis of wavelet-transformation are considered in article [Mazurova E.M., A.S. Bagrova, 2008]. Results of the spent experiment have shown, that compression of a kernel of Neumann's integral more than on 23% leads to accuracy loss at calculation of height anomaly of an order of two decimeters. Comparison of speed of calculations of a file in the size 32X32 a method of wavelet-transformation and a method of the Fast Fourier Transform (FFT) has shown considerable advantage of method of the FFT when we wish to receive heights of the geoid with centimeter accuracy. It occurs because at calculation of Neumann's integral by a wavelet-transformation method it is impossible to compress essentially a kernel of Neumann's integral without loss of accuracy of calculation of anomaly of height. Thus, it is impossible to recommend a method of wavelet-transformation for practical use at calculation of height anomaly in the central zone if only we do not wish to offer accuracy for economy of machine memory.

The paper [Mazurova E.M., Kuvekina N.A., Aleksashina E.V., 2008] presents the deduction of ellipsoid corrections in evaluating the disturbance potential from gravity disturbances. Spherical approximation makes the basis of almost all formulas of physical geodesy. From studies by D. Lelgemann, spherical approximation in application of Stokes' formula to determining the average height of the geoid all over the Earth leads to an error of the order of ±0.2 m. The modern accuracy of determining the anomalous potential of the gravitational field requires the flattening of the reference-ellipsoid to be taken into account. At present, satellite navigation technologies (GLONASS, GPS, etc.) allow gravity disturbances $g\delta$ to be calculated confidently, they being the function of the geodetic coordinates $B, L, H$ defined from satellite measurements.

The purpose of the work [Mazurova E.M., 2009] is development of an image Hartly for a kernel of the modified Vening-Meinez's integral. The concept of the deflection of the vertical is significant in physical geodesy. As the direction of the deflection of the vertical is defined by gravity, and evasion from them - gravity anomalies knowing these anomalies it is possible to find and deflection of the vertical. With development GNSS-technologies, it is possible to confidently define the gravity disturbance. Therefore authors can define the deflection of the vertical under a modified Vening-Meinez's formula. This integral is an integral of convolution and for its calculation in the central zone (radius to 10) it is effective to use two-dimensional fast Hartly transformation, using thus flat approximation. If the size of the central zone is to increase, it is necessary to change the integrated kernel. Fast Hartly transformation is especially effective, if the analytical image Hartly of an integrated kernel is known.

Spherical approximation makes the basis for a majority of formulas in physical geodesy. However, the present-day accuracy of determining the perturbing potential requires ellipsoidal approximation.

The papers [Mazurova E. M., Yurkina M. I., 2010a, Mazurova E. M., Yurkina M. I., 2010b] deals with constructing Green's function for an ellipsoidal Earth by a spherical harmonic expansion and using it for determining the perturbing potential. From the result obtained, the part corresponding to spherical approximation has been extracted. The remaining expansion is represented by a series whose terms decrease inversely proportional to the power number of the harmonic of the spherical harmonic expansion of the gravity force.

The Green function is known to depend only on the geometry of the surface where boundary values are set. Thus, it can be calculated irrespective of the gravity data completeness.

Any changes of gravity data do not have an effect on Green's function and they can be easily taken into account if it has already been computed. Such a solution method, therefore, can be useful in determining the perturbing potential of an ellipsoidal Earth.

Now, the central place in geodesy has been taken by satellite methods of co-ordinate determination using the signals of the GPS and GLONASS satellite navigation systems. Geodetic heights, however, computed from satellite measurements and normal heights determined by geometric leveling, exist independently from each other. The relationship between geodetic heights in a co-ordinate space and normal heights is executed by height anomalies which should be known with high accuracy. The problem of accurate determination of the gravity field transformants is caused by the fact that the classical methods of solving problems of physical geodesy are based on deriving and resolving some integrated equation. Continuous errorless values of gravity anomalies across the whole surface of the Earth are needed for that. But it is practically impossible to make a continuous gravimetric leveling, and the classical approach is not capable to produce an accurate answer without gravimetric leveling being completed. In practice, the initial information is discrete, with a lot of measuring errors, and known not across the whole surface of the Earth. Notice that Neumann's integral from which it is possible to calculate height anomaly is a convolution integral. Therefore, it is possible to use linear transformations for its calculation, such as the Fast Fourier Transform and wavelet-transform. In computing height anomaly, M.S.Molodensky's combined method is applied;

with it, numerical integration is conducted only within a certain ``limited area''. The influence of the distant field zone is taken into account by expending the gravity anomaly into series by spherical functions. The paper [Mazurova E.M., 2010] discusses the algorithm for calculating the height anomaly in the ``limited area'' in flat approximation. Neumann's integral calculations have been done on the basis of the fast discrete Wavelet-Transform. The results of the research into compressing the Neumann integral kernel that can be performed in calculating this integral by the Wavelet-Transform method are presented. An efficiency comparison of applying the Fast Fourier Transform and Fast Wavelet-Transform to computing the Earth's gravity field transformants has been made.

10th International Geoid School "The Determination and Use of the Geoid" was held on 28 June – 3 July 2010 at the State Research Center of the Russian Federation – Concern CSRI Elektropribor, JSC (Saint Petersburg, Russia). Young geodesists from Russian Federation, Germany, Canada, Turkey, USA as well as researchers of geophysics, geology and engineering took part in the school. Modern theories of geodetic gravimetry were enlightened and practical exercises of global model computation, lest square collocation and adjustment were passed.

To make a more exact mathematical model of the global atmosphere circulation and to improve the weather forecast one needs to take into account the Vertical Deflection Components (VDCs). The required accuracy for these tasks is 1–2″ now and twice better in future. The VDCs were calculated [Boyarsky E.A., et al., 2010] by integration of the free air gravity anomalies (taken from the Internet site of Scripps Institute of Oceanography) on the grid of $2' \times 2'$, an integration radius $R$ = 4,000 km. Also the approaches to calculate the far zone contribution are discussed. As the first test the results are compared with the VDCs obtained from the global spherical harmonics expansion of the order of 1,800. For the Mariana trench area, where the variation of $W$–$E$ component exceeds 50″, the standard deviation from Wenzel's model is of 2″. The other region for testing with a more dense grid $1' \times 1'$ is the Kane fracture zone; standard deviations are about 0″.6.

The MAGELLAN–2 package, developed at the Russian Academy of Sciences' Institute for Physics of the Earth (IPE) to use in a Windows environment, is a elaboration of the previous MAGELLAN version [Boyarsky E.A., Afanasieva L.V., 2010]. It is oriented toward modern gravimeters and the growing demand for accuracy and detail in measuring the gravity forces at sea and in the air.

The results of the measurements of gravity field at all the stations of gravity micronetwork at the BIPM are presented [Jiang Z., Becker M., Vitushkin L., 2008]. The BIPM is located on the slope of the hill in the park St Cloud in Sevres, France. This micronetwork consists of two sites with nine gravity stations in the buildings and two outdoor gravity stations. The difference of the values of free-fall acceleration at the gravity stations is about 9 milligal.

The technical protocol for the International Comparisons of Absolute Gravimeters developed according to rules for the Key Comparisons [Agostino G.D', Germak A., Vitushkin L.F., 2008]. The results of KCs provide the technical basis for the Mutual Recognition Arrangement of national measurement standards and of calibration and measurement certificates issued by national metrology institutes. The MRA is signed by more than sixty National Metrology Institutes over the world. The development of such Technical Protocol for ICAGs means that one more step to the increase of the confidence in gravity measurements was done. The details of the Technical Protocol are discussed.

The organization of ICAG-2005, measurement strategy, calculation and presentation of the results were described in a technical protocol pre-developed to the comparison. Nineteen absolute gravimeters carried out 96 series of measurements of free-fall acceleration g at the sites of the BIPM gravity network. The vertical gravity gradients were measured by relative gravimeters. For the first time the budgets of uncertainties were presented. The result of ICAG-2005 is in a good agreement with that obtained in ICAG-2001 [Vitushkin L., Jiang Z., et al, 2010].

A new method of the measurement of the gravitational constant using two free-moving discs and laser displacement interferometry is proposed [Vitushkin L., Wolf P., Vitouchkine A., 2007]. Such method can be realized in the labora-tory on the ground and on a spacecraft where the better accuracy in the measurements can be reached.

The basic ideas on the measuring standards in gravimetry are analyzed [L.F.Vitushkin, 2008]. It is shown that currently the primary reference measurement method of free fall acceleration is realized by the absolute ballistic gravimeter.

The successful results of the measurement of free fall acceleration using the absolute gravimeter FG5-108 with a compact frequency stabilized solid-state laser at 532 nm are presented in the paper [Vitushkin L., Orlov O., Nalivaev V., 2008].

The results of the measurement of the microseismics at the gravity stations of VNIIM at its Lomonosov branch are presented in [Krivtsov E., et al., 2008]. It is shown that besides of very quiet microseismics situation at this site a good damping is reached thanks to massive basement of 4000 tons.

The State Research Center of Russian Federation "Concern CSRI ELEKTROPRIBOR, JSC" and IAG with the support of Russian Foundation for Basic Research and Committee for Science and Higher Education of St Petersburg Government were organized two international IAG symposia "Terrestrial Gravimetry. Static and Mobile Measurements" - TGSMM-2007 and TGSMM-2010. Symposia were held in "Concern CSRI ELEKTROPRIBOR, JSC", Saint-Petersburg.

Main topics of the symposia were devoted to techniques, methods and results of terrestrial gravity observation, including the latest achievements in absolute gravity measurements using cold atomic gravity meters.

The X-th International Geoid School 'The Determination and Use of Geoid' has been organized by IGeS, the State Research Center of Russian Federation "Concern CSRI ELEKTROPRIBOR, JSC" and IAG with the support of Russian Foundation for Basic Research and Committee for Science and Higher Education of St Petersburg Government in the period from 28 June to 2 July 2010 at the CSRI ELEKTROPRIBOR in St Petersburg, Russian Federation. The well known scientists took part in the school as the invited lecturers.

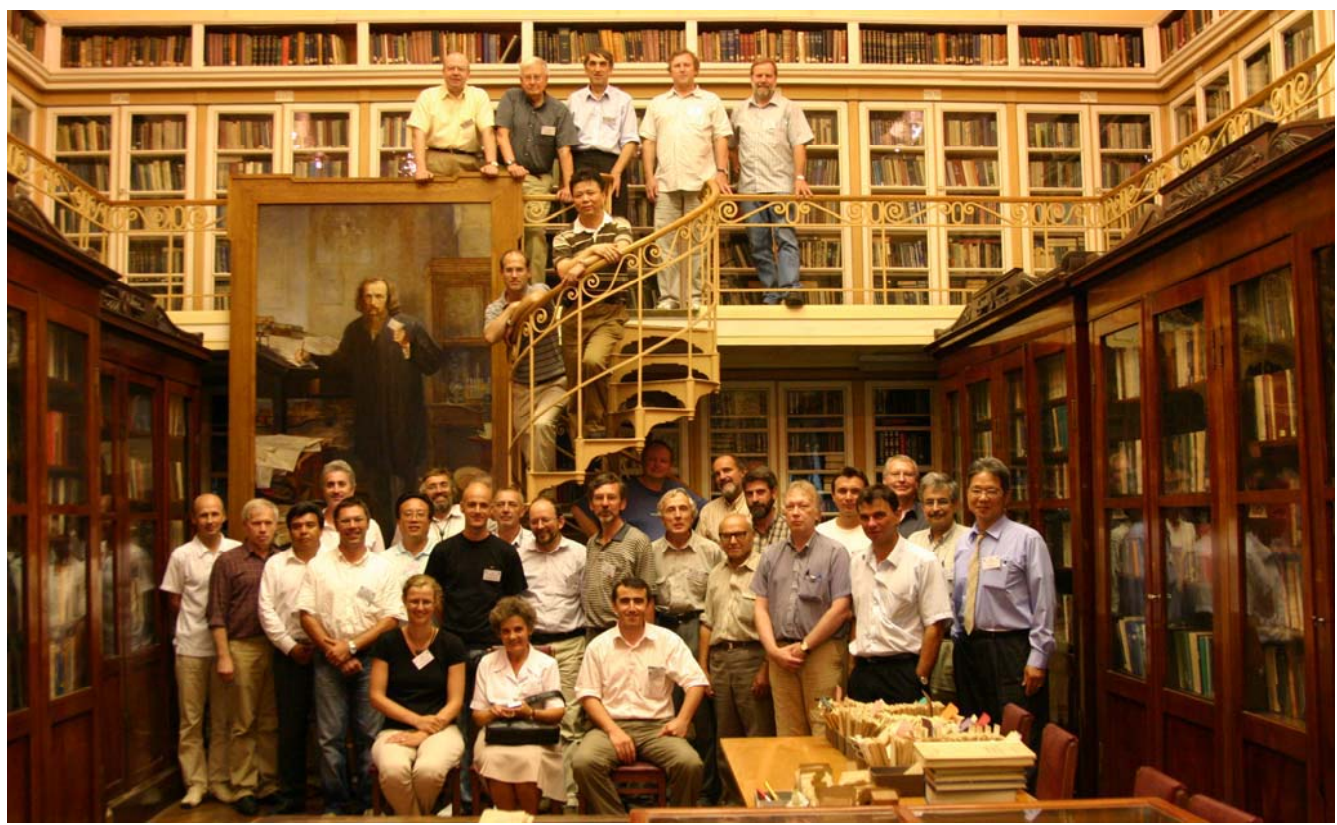

Fig.1 Participants of the 3rd Joint Meeting of CCM Working Group on Gravimetry andIAG Study Group on Comparisons of Absolute Gravimeters, library of VNIIM, StPetersburg, 24 August, 2010

The school was attended by 15 participants coming from 5 countries. The computers with O.S. were provided for all the students. Each computer worked with WinXP. The FORTRAN compilers and PHYTON interface have been installed to use FORTRAN programs that usually run under Unix systems. The students have been given Lecture Notes, IGeS CD with software and data for exercises, the GRAVSOFT manual, and a user guide for FFT exercises.

All the students received the certificates of successful graduation.

All-Russian D.I.Mendeleyev Research Institute for Metrology (Russian acronym VNIIM) proposed to organize the next International Comparison of Absolute Gravity Meters ICAG in 2017 on the reconstructed site for absolute gravity measurements in Lomonosov branch of VNIIM (Lomonosov is a little town in the vicinity of St Petersburg).

Now a new absolute ballistic gravimeter for the Primary State Standard of Russia in gravimetry as well as the document described the traceability chain for the measuring instruments in gravimetry is under development at VNIIM.

### 3. GEODYNAMICS
(Kaftan V.I., Steblov G.M., Tatevian S.K., Pevnev A.K.)

Nowadays a large amount of measurement information is collected in geosciences due to a great progress in technologies of observation. Space geodesy and geodynamics have collected unprecedented quantity of precise observation time series. The examples of this information can be the data of permanent GNSS, and VLBI, and gravity measurements or another geosciences observation. There are many long and high temporal resolution time series collected in modern geosciences' data bases. The large amount of the information creates conditions to find out new important regularities of the observed natural processes. This circumstance attracts many geodetic and geophysical institutes to analyze collected time series. Most of European geodetic institutes and organizations do research on the time series treatment. Some dominated periodical regularities are find in the observation and published in different scientific journals and issues. But the nature of them is not clear for today. An example the yearly periodicities in GPS observation can be interpreted as meteorological seasonal or tidal influences or GPS radio wave propagation. Each of them can be plausible explained but every source of the influence is not determined. There are some not understudied correlations between spectra of different observed processes. An example the spectra of solar activity and Caspian Sea water level have rather close correlation but the physical reasons are not known. These unclear correlations compel researchers to continue analysis of them for the reason of finding new natural regularities or uncorrected measurement errors. In the case of geodesy it is rather important to find the causes of revealed correlations. Such kind of research will bring new information for improving measurement accuracy and physical models.

The widely used procedures of analyzing a periodical behavior in time series have some disadvantages. The first is a not differentiated estimation of statistical confidence of revealed periodical components. The practice shows that a confidence of revealed periodical components varies significantly. An example yearly and daily oscillation harmonics usually have variations changing from hours to days. Standard techniques such as FFT or close to it not provide the detailed accuracy estimations for every of the confident components. For detailed analysis it is important to know RMS for every phase, amplitude and period estimations. This circumstance does not allow users of standard methods to simulate an accurate periodical model. The next weakness of typical techniques has not enough attention to phase determination. The phase analysis is performing very seldom and in terms of angular phase values. It does not bring clear information about phase shifts of near oscillation components. Temporal values of phase shifts are very important for the interrelation analysis for different observed processes.

For this task the original technique called as Sequential Analysis of Dominated Harmonics developed several years ago was used [Kaftan V.I., 2010]. This periodicity analysis technique based on the iteration least squares method was used for mathematical modeling of different natural processes. Time series of absolute gravity measurement, polar motion determination and GPS-hight measurement were used by the technique.

The method allows determining the trend component and the set of harmonics. The final model is a polyperiodical function that has the best fitting to the initial time series.

Two steps of computations were used: 1) estimation and removal of the trend (linear component), 2) finding one after another and removal of the high amplitude periodical components (harmonics), twice exceeded their standard errors. The proposed technique was used for forecasting of a behavior of important natural processes such as Caspian Sea level, solar cycle shape and pole coordinates. The predictions were statistically proved and demonstrated the effectiveness of the method.

Permanent GNSS measurements at sites of special geodetic networks provide high importance information for geodynamic purposes. They allow finding regularities of earth's surface movements and deformation.

The results of the permanent observation at the local GNSS network covering the San Andreas Fault were processed using the special technique of horizontal deformation computation and estimation of their uncertainty. The SOPAC archive data were used for computation of deformation characteristics from the beginning of 2004 to the end of 2007 with 10 day interval. Differences of the local network base line vectors were adjusted without any constrains. Invariant deformation characteristics and their standard errors were determined. The special ratio criterion was computed in order to determine the signal/noise information. The sample of more than hundred dispersion ratio values has shown the specific deformation intensity rise before the Parkfield earthquake occurrence. The rupture deformation and fault creep were determined rather surely as usual. The dispersion signal/noise criterion was rather low about a year before the earthquake but about four months later it begun steadily rise just to the main shock time. This specific deformation behaviour can be suggested as an evidence of a time geodetic precursor existence. In this case the situation of the earthquake forecast was quite convenient because the future rupture zone was expected about obviously and the large permanent GNSS network was specially constructed and monitored. The findings show that a strong earthquake moderate time forecast in the zones of expectation is possible.

Graphical representation of the results of Earth's movement and deformation determination are constructed by the authors of the paper [Kaftan V.I., et all, 2010]. Special video representation is recommended for long time series preliminary analysis to make easy deformation regularity research.

A few years ago the Global geodetic Observation System (GGOS) was initiated by the International Association of Geodesy and the International Union of Geodesy and Geophysics. The main task of the GGOS is the control of global changes by modern high technologies including global satellite navigation systems (GNSS) and computer techniques. For this reason the most of the modern global Earth observation results are related to the Electronic Geophysics Year themes.

The more or less spatially homogenous GNSS network was selected from the whole International Terrestrial Reference Frame (ITRF). Daily coordinate solutions were received from the Internet archive SOPAC for the interval of 1999-2005.5. Daily values of geocentric radius vectors were computed for all of 99 selected ITRF points. These data were used for the computer modeling of geometrical Earth shape behavior. The research [Kaftan V.I., Tsiba E.N., 2008, Kaftan V.I., Tsiba E.N., 2009abc] shows the interesting regularities in the performed geometric characteristic time series. The mean radius vector for the whole Earth demonstrates the rise tendency with the rate of 0.6 mm/yr. Geometrical Earth's ellipsoid semi major and semi minor axes grow both and have periodical oscillations with the amplitudes of 0.1-1.3 mm. The ratio of semi minor and semi major rise values shows the Earth's deflation decrease. The interesting feature is that the south hemisphere grows about three times faster than the north.

The results can be preliminary explained by both artificial and natural causes. The study produces the empirical data valuable for revealing some GNSS systematic errors and possible Earth's shape changes.

With a view to study variations of the geocenter movements, the sets of geocenter coordinates, estimated by the multi years sets of GPS and DORIS data processing, were analyzed. In order to estimate linear trend, amplitudes, periods and phases of geocenter variations a linear regression analysis has been applied with the use of least square method. The evaluated amplitudes of annual variations are of the order of 5-7 mm for X and Y components and 10-20 mm for Z component. Semi-annual amplitudes are also noticeable in all components. Secular trend in geocenter coordinates was found at the level of 2-3 mm.

The same time series of geocenter coordinates (X, Y, Z), have been examined with the use of so- called method of adaptive Dynamic Regression Modeling (DRM). Unlike the linear regression analyses, the DRM method envisages the further iterative, step by step, regression analyses of the non-random content of the detrended series, obtained after first harmonics removing, aiming to avoid errors, caused by noise residuals and inter correlation between estimated harmonics and to find out the additional regularities. As a result of DRM –method the original time series is approximated by the complex mathematical model, which contains trend, periodical components and parameters of the dynamic regression model. With this model a forecasting of weekly geocenter positions has been performed. The results are discussed in papers [Tatevian S.K., Kuzin S.P., 2009, Kuzin S.P., et al., 2010, Tatevian et al., 2010].

The paper [Kogan M. G., Steblov G. M., 2008] presents the vectors of rotation of 10 major lithospheric plates, estimated from continuous GPS observations at 192 globally distributed stations; 71 stations were selected as representing stable plate regions. All days for the period 1995.0–2007.0 were included in the analysis. In contrast to previous GPS plate models, proposed model is independent of international terrestrial reference frames (ITRF). The origin of formed plate consistent reference frame is the center of plate rotation (CP) rather than the center of mass of the entire Earth's system (CM) as in recent versions of ITRF. We estimate plate rotations and CP by minimizing the misfit between the horizontal velocities predicted by the plate model and the observed GPS velocities. If any version of ITRF is used as the reference frame, the drift of the ITRF origin relative to CP cannot be neglected in estimation of plate rotation vectors and plate residual station velocities. The model of the plate kinematics presented addresses the problem debated since the beginning of the space geodesy: how big are disagreements between the current plate motions and the motions averaged over several million years? The vectors of relative plate rotations estimated in the research with the published vectors from GPS and geologic models are compared. The integrity of individual plates as exhibited by plate residual station velocities is also discussed. For seven largest plates, the RMS value of plate residual station velocities in stable plate interiors is 0.5–0.9 mm/a; this value is an upper bound on deviation of real plates from infinite stiffness.

The present tectonics of Northeast Asia has been extensively investigated during the last 12 yr by using GPS techniques. Nevertheless, crustal velocity field of the southeast of Russia near the northeastern boundaries of the hypothesized Amurian microplate has not been defined yet. The GPS data collected between 1997 February and 2009 April at sites of the regional geodynamic network were used to estimate the recent geodynamic activity of this area. The calculated GPS velocities indicate almost internal (between network sites) and external (with respect to the Eurasian tectonic plate) stability of the investigated region. Authors [Shestakov N.V., et al., 2010] have not found clear evidences of any notable present-day tectonic activity of the Central Sikhote-Alin Fault as a whole. This fault is the main tectonic unit that determines the geological structure of the investigated region. The obtained results speak in favor of the existence of a few separate blocks and a more sophisticated structure of the proposed Amurian microplate in comparison with an indivisible plate approach.

Based on different-scale deformations monitoring in the Eurasian, North American, Pacific, Amur and Okhotsk lithosphere plates' junction area the problem of study of recent geodynamics of the Far Eastern region is discussed using contemporary space geodetic methods [Bikov V.G., et al., 2009]. A feasibility of the urgent need of formation of a unified observation network for deformations monitoring and seismological observations with the Far Eastern Branch, Russian Academy of Sciences, is shown. A brief review of the GPS observation results obtained within the framework of the Integration Project between FEB RAS and SB RAS in 2006–2008 is given. It is shown that application of space geo-detic data opens new opportunities for elaboration of the integral concept of the recent geodynamic activity in the East of Asia.

In 2006–2007, two great earthquakes ruptured the center of the Kuril subduction zone: first, the interplate thrust event, then the intraplate extensional event on the outer rise. The affected region was a seismic gap since 1915. Published patterns of slip differ for various seismic and tsunami inversions. The surface offsets that we measured with GPS on the Kuril Islands are sensitive to the total slip, including slow components beyond the seismic and tsunami band. The authors [Steblov G. M., et all, 2008a] invert coseismic offsets and show that the asperities, or regions of high slip, are spatially linked for both earthquakes; this pattern suggests (although does not prove) that the first event triggered the second. For the 2006 earthquake, the asperity is very shallow, probably because of the absence of an accretionary prism. For the 2007 earthquake, our modeling suggests that the rupture occurred in the bent Pacific lithosphere to a depth of 50 km.

At convergent plate boundaries, the accumulated elastic strain is released mostly in great shallow thrust earthquakes. Understanding which fraction of the elastic strain due to plate motion is released in earthquakes is important in assessing the seismogenic potential at convergent boundaries where most of great earthquakes occur. Conventionally, an estimate of the released fraction of the total elastic strain relies on the Kostrov summation of GCMT moment tensors. However, the GCMT moments of great earthquakes can be erroneous by a factor of 2 because of parameterization as a simple point source with triangular moment-rate function and because of trade-off between estimates of the moment and dip angle. Therefore, assessment of the current seismic potential caused by plate motion can be wrong. For the great 2006-2007 Kuril earthquakes, coseismic moments from inversion of GPS offsets are larger than GCMT moments by at least a factor of 1.5. The authors [Steblov G. M., et all, 2008b] attribute this disagreement to the simplified source model used in GCMT inversion. From kinematic GPS solution, a contribution of slow coseismic slip (captured by GPS but not by seismometers) was insignificant. Earlier, a similar disagreement between geodetic and GCMT moments was demonstrated for the great 2004 Sumatra earthquake.

Research on real plates and dubious microplates existence presented in [Kogan M. G., Steblov G. M., 2008]. From the onset of plate tectonics, the existence of most of the plates was never put in doubt, although the boundaries of some plates, like Africa, were later revised. There are however, two microplates in northeast Asia, the Amurian and Okhotsk, whose existence and the sense of rotation was revised several times. The rms value of plate-residual GPS velocities is 0.5-0.9 mm/yr for sets of stations representing the motion of the following plates: Antarctic, Australian, Eurasian, North American, Nubian, Pacific, and South American. This value can be regarded as an upper bound on deviation of real plates from infinite stiffness. The rms value of plate-residual GPS velocities is 1.2-1.8 mm/yr for the Indian, Nazca, and Somalian plates. Higher rms values for India and Nazca are attributed to the noisier data. The higher rms value for Somalia appears to arise from the distributed deformation to the east of the East African Rift; whether this statement is true can only be decided from observations of denser network in the future. From the analysis of plate-residual GPS velocities, the Canadian Arctic and northeastern Siberia belong to the North American plate. The detailed GPS survey on Sakhalin Island shows that the Sea of Okhotsk region also belongs to the North American plate while the region to the west of it belongs to the Eurasian plate. These results provide a constraint on the geometry of the North American plate and put in doubt the existence of smaller plates in northeast Asia.

North American plate (NAM), was earlier hypothesized judging from seismicity [Mackey et al., 1997]. Recent analysis of GPS observations in western Alaska, on islands in the Bering Sea interior, and on the Aleutians confirmed the existence of BER rotating clockwise around the pole in east Asia [Cross and Freymueller, 2008]. However, the confirmation of BER was lacking on the western margin. Here the authors [Kogan M.G., et all, 2009] augment the data base with the GPS survey of Chukotka performed at four epochs in 2004-2009 and analyze implications of velocities in Chukotka for the motion of BER. To strengthen reference to NAM in this region remote from the interior of NAM, the authors evaluated station velocities by the method adopted in the PBO processing: in each daily solution, we combined observations in Chukotka and at 25-30 well-determined global stations over a hemisphere centered on Chukotka. The data spanning five years showed two contrasting patterns of motion of stations in Chukotka. In general, whole Chukotka slowly moves relative to NAM at a speed not exceeding 3 mm/yr. While velocities over most of the region point to ESE, velocities over northeastern Chukotka point nearly to the south agreeing with the motion of St Lawrence Island and Seward Peninsula on Alaska, located to the east of Chukotka. The change in direction of motion over Chukotka indicates existence of a right-lateral fault that agrees with focal mechanisms of events over the seismic belt along the western margin of the Bering Sea. Movement of northeastern Chukotka confirms the sense and angular rate of rotation of BER as estimated by Cross and Freymueller from observations over the eastern part of BER. The slow motion of most of Chukotka to ESE indicates that this wedge-like extremity of NAM deforms being jammed between the Eurasian and Pacific plates, or, alternatively, it belongs to another rigid block located farther to the north.

The Kuril-Kamchatka subduction zone is the most mobile and seismically active region in Northeast Eurasia. The Kuril island arc is one of the few tectonically active regions, where until recently there had been no space geodetic network. The first GPS stations were installed on the Kamchatka Peninsula in 1997, and on the islands of the Kuril arc from Kamchatka to Hokkaido, in 2006. The collected geodetic data allowed to reveal the geometry of the interplate coupling along the whole Kuril_Kamchatka arc, and also to estimate the source parameters and their features for a number of major earthquakes in this area [G. M. Steblov, et all, 2010].

Russian and Finnish levelling networks were in 2002-2006 connected at 8 points along the Russian-Finnish boundary. We have now estimated contemporary vertical motion from joint processing of the repeated levellings in Northwest Russia and Finland. They span more than a century, and in large areas of the networks there are three campaigns. The authors [Mäkinen J., et all, 2008] compare the motion rates with velocities estimated from other measurement techniques and from geophysical models, and discuss their relationship with results from quaternary geology.

The yearly displacement velocity estimation technique for GNSS stations having the less irregular observation repetition (2-3 epochs) is proposed in [Shestakov N.V., et al., 2009]. It allows authors to reduce seasonal coordinate variation influence by taking into account apriory information parameters.

In the articles [Pevnev A.K., 2008-2010] the crisis reasons in a problem of the forecast of earthquakes are considered and the way of an exit from it is specified.

Mistake is motivated in interpreting the results of experiment targeted to checking the main position "The elastic-rebound theory of earthquakes" H.F.Reids. It is stated that American geodesists is allowed the blunder, concluding in that that they estimation of accuracy of results of measurements on undeformed base have used to measurements, run for deformed base. Exactly this and has brought them to conclusion about fallaciousness "The elastic-rebound theory of earthquakes", as was a source reason of crisis in decision of problem of forecast of earthquakes.

Natural geodynamic processes in the Earth crust become apparent as different types of movement. These movements can be regular when different strata move along fault lines on a local scale or along boundaries of tectonic plates on a continental scale. They can also be abrupt and have the form of earthquakes and other major seismic events. Both forms of movement are irreversible, and thus they can be called deformations.

Analysis of Earth crust motion and deformations can be performed by means of seismic soundings, strain-meter and tilt-meter readings, and geodetic methods which are the most significant ones. Geodesy plays a key role in crustal deformation studies by determining the temporal variations of the Earth shape and size at various spatial and time scales. Among modern geodetic methods space geodesy applications are of most importance. Among the latter Global Positioning Systems NAVSTAR-GPS and GLONASS should be pointed out.

Current research corresponds to an overview of certain GPS applications for Earth crust motion and deformation analysis and geody-

ГЕОДЕЗИЯ / GEODESYnamical and geophysical solutions [Krasnoperov R. (2010)].

The research on deformation in tectonically active and seismic zones is described in the papers [Prilepin M.T., 2007, Prilepin M.T., Baranova S.M., 2007].

The results of long-term observation of crustal deformations in the Northern Caucasus using long base laser strainmeter and GNSS network are presented in the paper [Milyukov,V.,et al.,2009]. An analysis of crustal strain laser interferometer records revealed a shallow magmatic chamber of Elbrus Volcano. First GNSS observation has shown the main movement velocity of 28 mm/yr in plane and uplift of the Northern Caucasus Ridge.

Within the last years Sternberg Astronomical Institute (SAI MSU) has established three new stationary GPS/GLONASS stations in the Northern Caucasus. The first (site code TRSK) is located in the Terskol International Astronomical Observatory near the Elbrus volcano, in the Kabardino-Balkaria Republic. It began to operate in 2005. The second one is located in the Solar Station of the Pulkovo Astronomical Observatory, Karatchay-Cherkessia Republic (site code KISL). This station has been in operation since 2006. The third one (site code VLAD) is located in Vladikavkaz (North Ossetia). This station is created together with the North Ossetia Scientific Center. The continuous GPS measurements began in 2008. These four stationary stations (including ZECK) form the base for the regional Northern Caucasus GPS network, which can be called the Northern Caucasus Deformation Array (NCDA).

At presence the repeated measurements of absolute gravity on the existing stations and the creation of the new stations are carried out by the ballistic gravity meters FG5 №110 and GABL. In 2007, after 13 years, the repeated absolute gravity measurements have been done at the stations "Zelenchukskaya" and SAI MSU". In 2008 two new stations, "Azau" (the Glaciology station of the MSU) and «Terskol», are established. In 2009 the repeated absolute gravity measurements were carried out at the stations "Azau" and "Zelenchukskaya". Tree new stations are created in the Northern Caucasus: "Nalchik", "Vadikavkaz" and "Ardon". Currently, the Northern Caucasus network for repeated measurements of the absolute gravity consists of seven stations – "Zelenchukskaya", "GAISH MSU", "Azau", "Terskol", "Nalchik", "Vadikavkaz" and "Ardon". Resent absolute gravity measurements were performed by joined team of SAI MSU and TsNIIGAiK.

The results of studies of local geodynamic processes in the scientific and educational base "Gornoe" of the State University of Land Use Planning are presented in the paper [Dokukin P.A., et al., 2009]. Studies were performed by the original method based on the analysis of the results of satellite geodetic measurements.

### 4. EARTH'S ROTATION
(Kaufman M.B., Malkin Z.M., Molodenskii S.M., Tolchel'nikova S.A.)

The problem of dependence of polar motion on precession-nutation model is discussed in several papers of M.I.Yurkina. It is actual since improvements in numerical, theoretical models are obtained by means of comparing spherical coordinates of points (projections of centers of bodies) – their calculated values with the observed ones. The former remains to be dependent on the motion of the Earth's axis in space and on the coordinates of the Earth's orbit. The efforts to improve the situation by means of solution of 3-body problem failed.

In answer to Euler's question "Qu'est ce que l'axe de la Terre" the author [Tolchel'nikova S.A., 2010] shows that the observed values which denote the motion of the Earth's axis in spherical coordinates (i.e. relatively to projections of stars ок quasars), are related to uncertain point inside the body of the Earth.

The aim of the paper [Tolchel'nikova S.A., 2010] is to better understanding between theorists and observers in their studies of the Earth's rotation.

Russian State Time, Frequency and Earth Rotation Service provides the official EOP data and time for use in scientific, technical and metrological works in Russia. The observations of GLONASS and GPS on 30 stations in Russia, and also the Russian and worldwide observations data of VLBI (35 stations) and SLR (20 stations) are used now. To these three series of EOP the data calculated in two other Russian analysis centers are added: IAA (VLBI, GPS and SLR series) and MCC (SLR). Joint processing of these 7 se-ries is carried out every day (the operational EOP data for the last day and the predicted values for 50 days). The EOP values are weekly refined and systematic errors of every individual series are corrected. The combined results become accessible on the VNIIFTRI server (ftp.imvp.ru) approximately at 6h UT daily [Kaufman M., Pasynok S., 2009].

Modern precise measuring techniques allow to define with high time resolution the Earth rotation parameters and a several of other geophysical processes (e.g. change of the angular atmospheric momentum). Authors of the paper [Kaufman M., Pasynok S., 2010] have performed an analysis of time shift between these processes in different frequency ranges. This investigation is of interest both in the theoretical plan, and in practical (e.g. for ERP prediction).

A comparison of the nutation theories IAU2000 and ERA2005 has been carried out on the efficiency of their use in a nutation angles prediction program [Pasynok S.L., 2008]. As a result, the used prediction procedure yielded similar results for ERA2005 and IAU2000 to within the uncertainties of the measurements.

According to recommendations of the IERS Workshop on EOP Combination and Prediction (Poland, Warsaw, 2009), the improvement of the model for FCN variation is necessary, particularly to provide greater accuracy for real-time and near real-time data processing. For achievement of this purpose it is necessary to consider a problem in a broad sense as construction of model of residual divergences of the IAU2000 nutation theory and results of practical VLBI measurements. The method based on correlation estimations and a method of the least squares (LSM) has been applied for the analysis of these divergences [Pasynok S., Kaufman M., 2010]. The basic results of the carried out research are:

1) The basic components of residual divergences theoretical and measured nutation angles are allocated, their periods and amplitudes are determined;
2) In a range of the periods from 400 to 500 day Z.Malkin and N.Miller's results of 2007 regarding presence in this range of two basic harmonics have been confirmed. However, values of the periods appeared a little bit distinguished (values of the periods of 405 and 462 day, instead of 410 and 452 day accordingly are received).
3) The assumption of presence of a harmonic with the period changing in time near to 433 day does not prove to be true results of the carried out analysis. Physical interpretation of the received results is discussed.

Though pure mathematical approximations like regression models, neural networks, etc. show good results in Earth rotation forecasting, dynamical modeling remains the only base for the physically meaningful prediction. It assumes the knowledge of cause-effect relations and physical model of the rotating Earth. Excitation reconstruction from the observed EOP is critical stage, needed for comparison with known causes, such as tidal forcing, AAM, OAM changes, and uncovering unknown ones. The authors [Zotov L., Kaufman M., 2009] demonstrate different approaches, which can be used here to avoid ill-conditionality and amplification of noises and present phase studies of the model and reconstructed excitations. They can be used for excitation forecast and EOP prediction through Kalman filtering.

In the paper [Malkin Z., 2008a], the author propose to use the scatter of celestial pole offset (CPO) series obtained from VLBI observations as a measure of the accuracy of the celestial reference frame (CRF) realizations. Several scatter indices (SIs) including those proposed for the first time are investigated. The first SI is based on analysis of residuals of CPO series w.r.t. Free Core Nutation (FCN) model. The second group of SIs includes Allan deviation and its extensions, which allow one to treat unequal and multidimensional observations. Application of these criteria to several radio source catalogues showed their ability to perform a preliminary assessment of the quality of radio source catalogues, and 2D Allan deviation estimate seems to be a most sensitive measure. However, in common case, the sensitivity of tested criteria is yet not sufficient to distinguish between radio source catalogues of the highest quality. Proposed extensions of Allan deviation, weighted and multidimensional, can be effectively used also for statistical analysis of astro-geodetic and other time series.

Results on studies which have been performed to investigate the impact of the cut-off elevation angle and elevation-dependent weight-ing on the EOP estimates and baseline length repeatability are presented in [Malkin Z., 2008b].

A new geometry index of Very Long Baseline Interferometry (VLBI) observing networks, the volume of network, is examined as an indicator of the errors in the Earth orientation parameters (EOP) obtained from VLBI observations [Malkin Z., 2009]. It has been shown that both EOP precision and accuracy can be well described by the power law $\sigma = aV^c$ in a wide range of the network size from domestic to global VLBI networks. In other words, as the network volume grows, the EOP errors become smaller following a power law. This should be taken into account for a proper comparison of EOP estimates obtained from different VLBI networks. Thus performing correct EOP comparison allows us to accurately investigate finer factors affecting the EOP errors. In particular, it was found that the dependence of the EOP precision and accuracy on the recording data rate can also be described by a power law. One important conclusion is that the EOP accuracy depends primarily on the network geometry and to lesser extent on other factors, such as recording mode and data rate and scheduling parameters, whereas these factors have stronger impact on the EOP precision.

During close angular approaches of solar system planets to astrometric radio sources, the apparent positions of these sources shift due to relativistic effects and, thus, these events may be used for testing the theory of general relativity; this fact was successfully demonstrated in the experiments on the measurements of radio source position shifts during the approaches of Jupiter carried out in 1988 and 2002. An analysis, performed within the frames of the present work, showed that when a source is observed near a planet's disk edge, i.e., practically in the case of occultation, the current experimental accuracy makes it possible to measure the relativistic effects for all planets. However, radio occultations are fairly rare events. At the same time, only Jupiter and Saturn provide noticeable relativistic effects approaching the radio sources at angular distances of about a few planet radii. The analysis [Malkin Z., et al., 2009] resulted in the creation of a catalog of forthcoming occultations and approaches of planets to astrometric radio sources for the time period of 2008–2050, which can be used for planning experiments on testing gravity theories and other purposes. For all events included in the catalog, the main relativistic effects are calculated both for ground-based and space (Earth–Moon) interferometer baselines.

A new method is developed for prediction of UT1 [Tissen V., et al., 2010]. The method is based on construction of a general polyharmonic model of the Earth rotation parameters variations using all the data available for the last 80-100 years, and modified autoregression technique. A rigorous comparison of UT1 predictions computed at SNIIM with the prediction computed by IERS (USNO) in 2008-2009 has shown that proposed method provides better accuracy both for ultra-short and long term predictions. VLBI observations carried out by global networks provide the most accurate values of the precession–nutation angles determining the position of the celestial pole; as a rule, these results become available two to four weeks after the observations. Therefore, numerous applications, such as satellite navigation systems, operational determination of Universal Time, and space navigation, use predictions of the coordinates of the celestial pole. In connection with this, the accuracy of predictions of the precession–nutation angles based on observational data obtained over the last three years is analyzed for the first time, using three empiric nutation models—namely, those developed at the US Naval Observatory, the Paris Observatory, and the Pulkovo Observatory. This analysis shows that the last model has the best of accuracy in predicting the coordinates of the celestial pole. The rms error for a one-month prediction proposed by this model is below 100 microarcsecond [Malkin Z. M., 2010].

Astrometric observations of the radio source occultations by solar system bodies may be of large interest for testing gravity theories, dynamical astronomy, and planetary physics. In the paper [L'vov V., et al., 2010], authors present an updated list of the occultations of astrometric radio sources by planets expected in the nearest years. Such events, like the solar eclipses, generally speaking, can be only observed in a limited region. The map of the shadow path is provided for the events occurred in regions with several VLBI stations and hence the most interesting for radio astronomy experiments.

Investigations of the anomalies in the Earth rotation, in particular, the polar motion components, play an important role in our understanding of the processes that drive changes in the Earth's surface, interior, atmosphere, and ocean. The paper [Malkin Z., Miller N., 2010] is primarily aimed at investigation of the Chandler wobble (CW) at the whole available 163-year interval to search for the major CW amplitude and phase variations. First, the CW signal was extracted from the IERS (International Earth Rotation and Reference Systems Service) Pole coordinates time series using two digital filters: the singular spectrum analysis and Fourier transform. The CW amplitude and phase variations were examined by means of the wavelet transform and Hilbert transform. Results of the analysis have shown that, besides the well-known CW phase jump in the 1920s, two other large phase jumps have been found in the 1850s and 2000s. As in the 1920s, these phase jumps occurred contemporarily with a sharp decrease in the CW amplitude.

In the papers [Molodenskii, 2010, 2011] the ambiguity in the inverse problem of retrieval of the mechanical parameters of the Earth's shell and core from the set of data on the velocities and of longitudinal and transverse seismic body waves, the frequencies and quality factors of free oscillations, and the amplitudes and phases of forced nutation is considered. The numerical experiments show that the inverse problem of simultaneous retrieval of the density profile in the mantle–liquid core system and the mechanical quality factor of the mantle (if the total mass M and the total mean moment of inertia I of the Earth, and Vp and Vs are constant at all depths) has most unstable solutions. An example of depth distributions of density and quality factors which are alternative to the well-known PREM model is given. In these distributions, the values of M and I and the velocities Vp and Vs at all depths for the period of oscillations T = 1 s exactly coincide with their counterparts yielded by PREM model (T = 1 s); however, the maximum deviations of the density and quality factors profiles from those in the PREM model are about 3% and 40%, respectively; the mass and the moment of inertia of the liquid core are smaller than those for the PREM model by 0.75% and 0.63%, respectively. In this model, the root mean square (rms) deviations of all the measured values of free oscillations frequencies $\sigma_i$ and $Q_i$ from their values predicted by theory are half to third the corresponding values in the PREM model; the values of rms for natural frequencies of the fundamental tone and overtones of radial oscillations, the fundamental tones of toroidal oscillations, and the fundamental tones of spheroidal oscillations, which are measured with the highest relative accuracy, are smaller by a factor of 30, 6.6, and 2 than those in the PREM model, respectively.

Such a large ambiguity in the solution of the inverse problem indicates that the current models of the depth distribution of density have relatively low accuracy, and the models of the depth distribution of the Q- factors in the mantle are extremely unreliable.

It is shown that the ambiguity in the models of depth distribution of density considerably decreases after the new data on the amplitudes and phases of the forced nutation of the Earth are taken into account. Using the same data, one may also refine by several times the recent estimates of the creep function for the lower mantle within a wide interval of periods ranging from a second to a day.

In the papers [Zharkov V.N., Gudkova T.V., Molodenskii S.M., 2009], [Molodenskii S.M., Zharkov V.N., Gudkova T.V., 2009] the theory of forced nutation of the tri-axial planet is constructed. The results are applied to the Mars' forced nutation.

## 5. POSITIONING AND APPLICATIONS
(Kaftan V.I.)

Results of the research on satellite laser range distance errors and techniques are presented in the papers [Prilepin M.T., et al., 2008, 2009].

Research on geodetic measurements and monitoring of large bridge construction using modern GNSS and electronic technologies carried out by the authors of publications [Brin' M.Ya., et al., 2007, 2009, Nikitchin A.A., 2009, Sergeev O.P., et al., 2008].

Laser scanning techniques are widely used in engineer application of geodesy. Some theoretical and practical researches are presented in the papers [Kanashin N.V., Kougiya V.A., 2007, 2009].

Series of works devoted to a designee of modern gravimetric equipment and techniques are presented in publications listed below.

The article [Churilov I.D., Dokukin P.A., 2010] describes an operational experience of the use of positioning system at pipeline inspection in the South China Sea with used ROV (Remotely Operated Vehicle). The description of the used positioning method, devices and software is presented.

Results of the study of an influence of discrete digital device electronic total stations on the accuracy of observation results of point sources of emission are described in the paper [Baranov V.N., Bragin A.A., 2009].

## 6. THEORY AND COMPUTATION
(Gerasimenko M.D., Neyman Yu.M., Tolchel'nikova S.A., Shestakov N.V.)

The relations between geodesy and gravimetry are discussed in the monograph [Gravimetry and Geodesy, 2010, 570 p. In Russian] which is dedicated to the 90th birthday of Vsevolod V.Brovar, an outstanding Russian scientist in the domain of geodetic gravimetry and geodesy, and to the centenary of Mikhail S.Molodensky, a Russian geodesist and geophysicist, the founder of the contemporary Earth figure theory, who made geodesy free from hypotheses of the Earth internal structure and has transformed geodesy into an exact science.

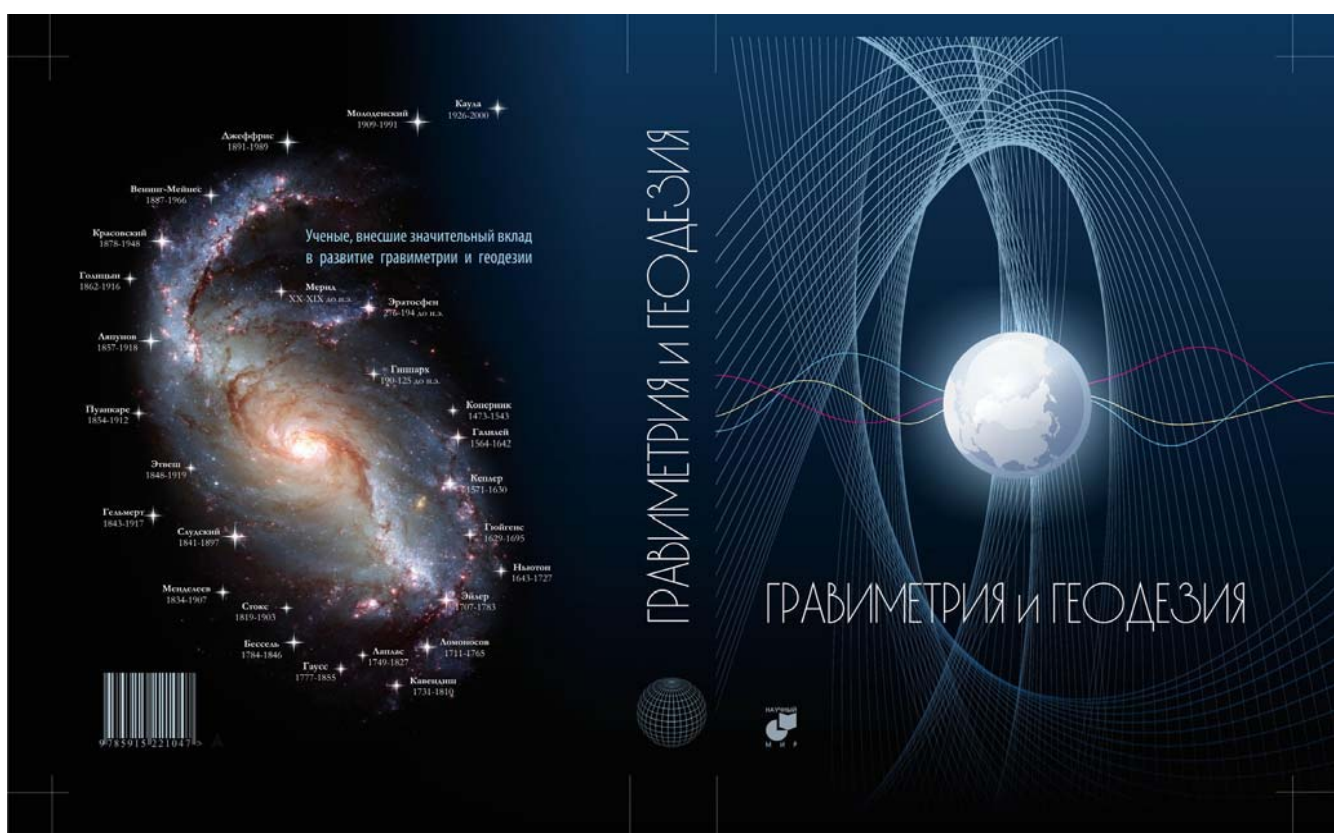

Fig. 10 Covering of the book "Gravimetry and Geodesy"

The purpose of the book is to show fruitful interacted influences of gravimetry and geodesy, evolutions of their goals and tasks; to sum up the main theoretical, experimental and production work done at the verge of the 2nd an 3rd millennia in the field of geodetic gravimetry (or physical geodesy), the transition from particular to system solutions of geodesy and gravimetry problems; to show the prospects for gravimetry development in the interests of geodesy.

The book is intended for the attention of geodesists, gravimetrists, astronomers, geophysicists, builders, navigators, ballisticians, designers, organizers of science and of all those who are interested in the development of geodesy and gravimetry [Gravimetry and Geodesy, 2010].

The determination of the motions of the Earth's pole and the places of observations in XX century is reviewed in the paper [Tolchel'nikova S.A., 2010]. Information is given about the his-tory of the spherical coordinate systems used by different International services monitoring the Earth rotation.

The principal possibility is shown to connect the mean coordinates of reference objects (stars, quasars) to the constant coordinates- λ, φ- of the one equatorial observatory which permits to exclude the usage of numerical (theoretical) models for precession and nutation to the obtained long periodic and secular motions under discussion. At the first step projections of mean polar places to the celestial sphere are considered.

In [Tolchel'nikova S.A., 2008] the transition is considered from studies of the motions of projections of objects in spherical coordinate systems to those of bodies in triads. The views of E.Mach and A.Einstein admitting only relative motion are not heuristic. As an example, the SRT formulae for stellar aberration are shown to be not adequate for the great velocity of an observer. Instead of their "improvement" using GRT corrections for space curvature, it is necessary to take into account the curvature of celestial sphere (a surface) in the case when the velocity of observer is not less than 0.01c and/or when the error of observation is less than 0.5ms of arc. For the latter case the formulae of stellar aberration were derived in Izvestija GAO (Pulkovo) № 213, 1998 and № 214, 2000.

In [Tolchel'nikova S.A., 2009] two periods are considered in the development of astronomy and geodesy which have been a single science, and were divided into two close cooperating departments in XIX century. After technical rearmament in both branches of science and the latest "revolution in astrometry" the differentiation in the two fields of activity has increased. Under discussion is the newest terminology, frequent changes in theoretical models implemented by IERS and reverberated in IAU Resolutions, the influence of cardinal changes in methods of establishing reference systems ICRF and ICRS on the studies of long periodic and secular problems in Earth's rotation. The attention is paid to the actual idea of establishing International Earth's Coordinate System advanced by the scientists of TsNIIGAiK.

In order to improve cooperation between scientists employing different methods for studies of motion, the method is considered used by Copernicus for simultaneous determination of secular and periodic components of the Earth's orbital motion and rotation of its axis from the values of coordinates fixed in observations during 1500 years. In modern practice secular and periodic components are determined separately.

For determination of heliocentric motions which gave start to the origin of dynamics, Copernicus had to summarize the observed changes of spherical coordinates (hence leading role was played by inductive logic) whereas the thought of modern scientists is moving in opposite direction since they use numerical theoretical models for interpretation of observed changes of coordinates. The development of inductive method is required for synthesis of information received from observations in astronomy, geodesy and so on, particularly for solution of the actual problem of separate determination of rotations of two celestial circles (equator and ecliptic) from analysis of observed coordinates' values [Tolchel'nikova S.A., 2010a]. The difficulties are shown in solution of this problem by means of mechanical methods and the simplified method employed in XVII century.

After Newton, Euler developed the dynamics of rigid bodies in general and in application to several planets. Attention is paid to Euler's studies of rotation of the Earth's axes in space (relative to the celestial coordinate system) and to his prediction of polar motion [Yurkina M.I., Tolchel'nikova S.A., 2009]. The influence of polar motion to the fluctuations of observed latitudes was found in more than hundred years after Euler's prediction. In XX century different International services have been organized for monitoring latitudes φ and longitudes λ of the observatories. Deviations from Euler's value for the period of free polar motion are under discussion up to now. Euler's ideas are shown to be helpful to modern geodesists and astronomers for consolidation of their efforts.

Short information [Tolchel'nikova S.A., Yurkina M.I., 2007] is given about the great influence of Euler's heritage and the International scientific conference "Leonard Euler and modern science" which took place in May 2007 in St.-Petersburg, where the authors presented the paper "Determination of the pole of the Earth's rotation", published in the volume "Leonard Euler and modern science", СПб., РАН, 2007, с.305-311.

Variation method of physical geodesy and collocation represented by Yu. Neyman and V. Bywshev described in the paper [Neyman Yu.M. & Bywshev V.A., 2010]. Functional approach to physical geodesy problems is described. Tractability of any geodetic measurement subject as a functional on geopotential is a guide for it. The geopotential is treated as element of a Hilbert space with reproducing kernel. Practical choosing of the kernel is founded on the well known isomorphism between reproducing kernel Hilbert spaces and covariance theory of random functions. The publication is an essential supplemented version of lecture notes published by MIIGAiK in 1985 - 1986.

It is often fruitful to use inhomogeneous information variable by accuracy and measurement unit for a perfect estimation of latent parameters. Ignoring of a physical value in this usage can produce not correct conclusions. The brief review of the iteration procedure of latent parameter estimation using inhomogeneous measurement results in the frame of the classic least square method presented in [Neyman Yu.M., 2008].

The outcome of the different geoid computation methods, based on gravity, using methods such as FFT, collocation or Stokes integration, all give the *gravimetric geoid* - which in principle refers to a global reference system, i.e. global center of mass, average zero-potential surface etc. Such a geoid may be substantially offset from the apparent geoid heights determined from GPS and leveling. The reason for the difference is mainly the assumption of zero-level: levelling zero refers to local or regional mean sea-level, which is different from the global zero vertical datum due to the sea-surface topography. Since isolated territories are interested in using GPS to determine heights in a local vertical datum, to be consistent with existing levelling, there is a need to taylor the gravimetric geoid to the local level. The paper [Neyman Yu., Pham Hoang Lan, 2010] gives a survey how it can be done.

The equations of connections and corrections for the SST measured values are derived and proposed to Earth's gravity parameters improving in the paper [Neyman Y.M., et al, 2008].

The main information needed for handling with files in European Space Agensy standard are described. Examples of output GOCE mission data are given in the article [Neyman Y.M., Hoziaichikov A.A., 2010].

A simple derive of the energy conservation of the Earth - satellite system is shown without using of the analytical mechanic terms. Gravity potential and disturbing gravitation modeling by the satellite state vectors are written [Neyman Y.M., Sugaipova L.S., 2011].

The theory of Fourier series sampling theorem is applied to calculation of covariance between point and mean values of various transforms of geopotential. The corresponding numerical results for the second derivatives of the disturbing potential are obtained to use for primary treatment of satellite gradiometry data [Sugaipova L.S., 2010].

The main results of theoretical and experimental research on optimal observation design and rigorous mathematical treatment of geodetic measurement for geodynamical purpose carried out in the Institute of Applied Mathematics of the Far East Division of the Russian Academy of Sciences during the last 20 years [Gerasimenko M.D., Kolomiets A.G., 2008].

The problem of the gross error diagnostics based upon the analysis of least-squares residuals has been carried out in [Gerasimenko M.D., 2008]. The general formula for calculation of the square root of the diagonal elements of covariance matrix of the residuals is given. This formula allows calculating the critical value of residuals, including measured directions even under terms when the position-finding angles are excluded from the system of normal equations.

Recommendations to take into account a physical correlation in mathematical treatment of geodetic measurements proposed in the article [Gerasimenko M.D., Shestakov N.V., 2008].

An explicit relationship between accuracy of estimated parameter of deformation model, weights of optimized GPS observations and degree of their correlation was obtained and investigated for the first time [Shestakov N.V., Gerasimenko M.D., 2009a]. It is shown that taking into account correlation of GPS measurements may lead to increase of formal precision of estimated parameters. Thus, we don't recommend to involve correlation of GPS measurements into optimization process without knowledge of real values of correlation coefficients.

The elasticity theory basic aspects necessary for deformation description and analysis, simple deformation models, deformation GNSS network and measurement characteristics, geodetic network optimal design problems are presented in [Shestakov N.V., Gerasimenko M.D., 2009b, Shestakov N.V., et al., 2008]. Deformation GNSS network optimal design and most informative observation selection algorithms are proposed. Correlated geodetic measurement optimization problems are considered and analyzed. The major theoretical conclusions and practical developments are illustrated by numerical examples.

As it is shown by V.Krilov (2008a) the main disadvantage of the Encke method for numerical integration of celestial body movements, related to increasing computation operations in the account of two orbit integration, can be eliminated with the use of regularized equations. The example of integration of movement equations of two asteroids shows that the method Encke usage for regularized equations provides coordinate accuracy at the level of $6*10-7$ of astronomical unit at the 400 day interval with about 230 steps of the integration.

Using three Laplace integrals and integral of energy, regularized equations of undisturbed and disturbed movement of an astronomical object are received in the research [Krilov V.I., 2008b]. It is shown that undisturbed astronomical object coordinates can be presented as a function of initial conditions and a new independent variable. The example of integration of movement equations of the 1566 Icarus asteroid, having large orbital eccentricity, shows that the regularization provides coordinate accuracy at the level of $5*10-7$ of astronomical unit at the 400 day interval with about 430 steps of the integration.

The book "Spherical functions in a geodesy" written by Zbynek Nadenik the researcher from the Czech Republic were interpreted to Russian and edited as a student book [Nadenik Z., 2010].